\documentclass[useAMS,usenatbib]{mn2e}

\usepackage{amsmath}
\usepackage{times}
\usepackage{graphicx}
\usepackage{booktabs} 
\usepackage{dcolumn}
\usepackage{fixltx2e} 
\usepackage{relsize} 
\newcommand{\bvec}[1]{\ensuremath{\mbox{\boldmath$\displaystyle#1$}}}

\def\muas{$\mu$as}

\def\vecp{{\bvec p}}
\def\vecphi{{\bvec \phi}}

\def\logg{\log g}
\def\feh{\rm [Fe/H]}
\def\A0{A_0}
\def\AG{A_{\rm G}}
\def\MG{M_{\rm G}}
\def\R0{R_0}\def\teff{T_{\rm eff}}
\def\parallax{\varpi}

\def\uas{$\mu as$}

\def\ltsim{\:{_<\atop{^\sim}}\:}

\def\ilium{\textsc{ilium}}

\title[Stellar parameter estimation with Gaia]{The expected performance of stellar parametrization with Gaia spectrophotometry}
\author[C.\ Liu et al.]{C.\ Liu$^1$,  C.A.L.\ Bailer-Jones$^1$, R.\ Sordo$^2$, A.\ Vallenari$^2$, R.\ Borrachero$^3$, X.\ Luri$^3$, P.\ Sartoretti$^4$\\
$^1$Max Planck Institute for Astronomy, K\"onigstuhl 17, Heidelberg, 69117, Germany\\
$^2$INAF, Osservatorio Astronomico di Padova, Vicolo Osservatorio 5, Padova, Italy\\
$^3$Dept. Astronomia i Meteorologia ICCUB-IEEC, Mart\'i i Franqu\`es 1, Barcelona, Spain\\
$^4$GEPI, Observatoire de Paris, CNRS, Univ. Paris Diderot, Place Jules Janssen, 92190 Meudon, France\\
}

\begin{document}

\date{Submitted 30 May 2012; resubmitted 25 July 2012; accepted 25 July 2012; minor corrections 27 August 2012}

\maketitle


\begin{abstract} 
Gaia will obtain astrometry and spectrophotometry for essentially all sources in the sky down to a broad band magnitude limit of G=20, an expected yield of $10^9$ stars.  Its main scientific objective is to reveal the formation and evolution of our Galaxy through chemo-dynamical analysis. In addition to inferring positions, parallaxes and proper motions from the astrometry, we must also infer the astrophysical parameters of the stars from the spectrophotometry, the BP/RP spectrum. Here we investigate the performance of three different algorithms (SVM, \ilium, Aeneas) for estimating the effective temperature, line-of-sight interstellar extinction, metallicity and surface gravity of A--M stars over a wide range of these parameters and over the full magnitude range Gaia will observe (G=6--20\,mag). One of the algorithms, Aeneas, infers the posterior probability density function over all parameters, and can optionally take into account the parallax and the Hertzsprung--Russell diagram to improve the estimates. For all algorithms the accuracy of estimation depends on G and on the value of the parameters themselves, so a broad summary of performance is only approximate. For stars at G=15 with less than two magnitudes extinction, we expect to be able to estimate $\teff$ to within 1\%, $\logg$ to 0.1--0.2\,dex, and \feh\ (for FGKM stars) to 0.1--0.2\,dex, just using the BP/RP spectrum (mean absolute error statistics are quoted).  Performance degrades at larger extinctions, but not always by a large amount.  Extinction can be estimated to an accuracy of 0.05--0.2\,mag for stars across the full parameter range with a priori unknown extinction between 0 and 10\,mag.  Performance degrades at fainter magnitudes, but even at G=19 we can estimate $\logg$ to better than 0.2\,dex for all spectral types, and $\feh$ to within 0.35\,dex for FGKM stars, for extinctions below 1\,mag.
\end{abstract}


\section{Introduction}\label{sect:intro}

The Gaia satellite will undertake the most detailed and accurate astrometric census of stars in our Galaxy ever attempted. During the course of its five year mission, Gaia will measure the positions, parallaxes and proper motions of some $10^9$ stars -- essentially all point sources in the sky brighter than a visual magnitude of 20 -- with an expected parallax accuracy of 10--25\,\muas\ at 15$^{th}$ magnitude (\citealp{debruijne12}). In addition to these five components of phase space, Gaia will also measure the sixth -- radial velocity -- for stars brighter than about 17$^{th}$ magnitude. From these data we can construct a detailed spatial and kinematic map of a statistically significant part of our Galaxy, which will form the basis for understanding its composition, formation and evolution (e.g.\  \citealp{perryman01}, \citealp{turon05}, \citealp{brown05}, \citealp{lindegren08}, \citealp{cbj09}).\footnote{\tt http://www.rssd.esa.int/Gaia}

However, kinematic information on anonymous stars is of limited use, and for this purpose Gaia is equipped with two low resolution prism spectrophotometers which together provide the spectral energy distribution (SED) of all targets from 330--1050\,nm, with a resolution varying between 13 and 85.
The main purpose of these spectrophotometers (named BP and RP, for ``blue photometer'' and ``red photometer'') is to classify the objects (into star, galaxy, quasar etc.) and to determine the stars' physical properties (as well as to provide a chromaticity correction for the astrometric data analysis).  Reliable knowledge of these properties is central to achieving an understanding of the stellar population of the Galaxy. While the most fundamental properties of a star are its mass, age and composition, these must be inferred via models, and, for the majority of stars, via the atmospheric parameters of (primarily) effective temperature, surface gravity and metallicity. The interstellar extinction must also be estimated, ideally star-by-star, as only then can we infer the absolute magnitude (and hence luminosity) from the apparent magnitude and parallax.

In this article we describe the design and current performance of the algorithms which will be used to estimate these four stellar astrophysical parameters from the Gaia data ($\teff$, $\A0$, \feh, $\logg$).
These algorithms are part of the Gaia data processing system currently under development. This system does the complete reduction of the data -- from telemetry stream to final catalogue -- and is being developed by (and will be operated by) the Gaia Data Processing and Analysis Consortium (DPAC; \citealp{aoresponse}, \citealp{mignard08}).

The algorithms described in this article are collectively referred to as {\em GSP-Phot}, the {\em General Stellar Parametrizer} based on (spectro)Photometry (i.e.\ BP/RP). This is part of the {\em Apsis} pipeline, that part of the DPAC system which performs the object classification and astrophysical parameter estimation.
GSP-Phot will be applied to every one of the approximately $10^9$ objects which Gaia will observe (the vast majority of which will be stars).
Apsis also includes a module to perform discrete classification (e.g.\ \citealp{cbj08}), quasar and galaxy spectral parameter estimation (e.g.\ \citealp{vivi12}), as well modules for classifying specific types of stars such as physical binaries or emission line stars (e.g.\ \citealp{bloome11}). A separate module estimates stellar parameters using spectra from Gaia's radial velocity spectrograph (RVS), although 
just for the brighter stars ($G\ltsim14$) (e.g.\ \citealp{recio06}). Although this instrument has a much higher resolution (11\,500) than BP/RP, it  also has a much narrower wavelength coverage (847--874\,nm) and achieves lower signal-to-noise ratios (SNRs) than BP/RP for a given magnitude (see \citealp{katz04} for a description of this instrument). A final pair of algorithms performs unsupervised cluster analyses and outlier analysis.

There exist several different algorithms in the astronomical literature which have been used to estimate stellar parameters from (spectro)photometric data. Without trying to be complete, these include physical line-based analyses (\citealp{beers99}, \citealp{heiter03}), nearest neighbours (often $\chi^2$) template-based methods (\citealp{wilhelm99}, \citealp{ap06}, \citealp{belikov08}, \citealp{straizys11}), neural networks or kernel regression (\citealp{cbj98}, \citealp{wjd04}, \citealp{zhang06}, \citealp{rf07}),
methods based on forward modelling schemes (\citealp{cayrel91}, \citealp{takeda95}, \citealp{koleva09}, \citealp{cbj10}) and Bayesian methods (\citealp{hill09}, \citealp{mortlock09}, \citealp{cbj11}). While these have generally been quite successful, many have been applied either to small data sets, to data sets with a limited and known range of stellar parameters, or have only attempted to estimate one or two parameters. The Gaia case is different, because by design -- an all-sky, deep, magnitude-limited and confusion-limited survey -- it will observe a broad, multi-parameter space. It is partly for this reason that Apsis contains, in addition to the ``general'' algorithms described here, algorithms customized for specific types of stars (taking their inputs from the inevitably coarser parameter estimates obtained by GSP-Phot).

As described in the next section, GSP-Phot is composed of three algorithms which estimate parameters separately from one another. The logic behind this is the collective experience in machine learning that there is rarely one  algorithm which is the best across all parameter space. The results in this article confirm this. Working with the Gaia data will be a learning experience, so it would be unwise to rely on just one algorithm. (The SDSS/SEGUE survey also used multiple algorithms; \citealp{lee08}.)

Two of the algorithms (SVM and \ilium) use only the BP/RP spectrum (examples of the data are shown in section~\ref{sect:data}). The third algorithm (Aeneas) can additionally use the Gaia-measured parallax and apparent magnitude (plus uncertainties). Both \ilium\ and Aeneas provide specific uncertainty estimates on the parameters (with Aeneas providing a complete posterior probability distribution over the parameters). Taking also into account the other algorithms in Apsis, most stars will receive multiple estimates of each parameter. All of these will be reported in the final Gaia catalogue, along with a ``best estimate'' in each case. (The scheme for producing this has not yet been finalized.)  Inference always depends on assumptions in addition to the data, so the advanced user will need to take these into account when deciding which estimates to use. Ultimately the GSP-Phot software could be made available, in order to permit users to change the assumptions (e.g.\ input libraries) and to reparametrize some or all of the Gaia data (which will also be publicly released).

Gaia is scheduled for launch in late 2013 with observations continuing until 2019.  While much of the Gaia data processing software is already in place, it will be continually improved during the mission as we learn more about the data quality and the instruments' behaviour (and how these evolve over five years in the Earth--Sun L2 environment). This article describes the GSP-Phot algorithms and their performance on simulated data as of early 2012. Although significant effort has been taken to correctly simulate the observed Galaxy, the instruments and the various sources of noise, real data are inevitably different and more complicated.  Indeed, a significant part of our remaining work is to develop and apply a suitable calibration of the algorithms and/or the input spectral libraries in order to accommodate mismatch between real and simulated data. 
Thus true performance may differ. Caveat lector! 

After outlining the algorithms in section~\ref{sect:meth}, section~\ref{sect:data} describes the simulated data used to train and test the algorithms. In section~\ref{sect:res1} we present the results, and in section~\ref{sect:disc} we compare the results from the different algorithms with each other and with expectations, and look at specific regions of parameter space (the ``science cases''). We conclude in section~\ref{sect:summ}. Electronic tables of the full results are available in the online version of this article, and are described in the appendix. The reader can use these to perform further analyses or make additional performance predictions.
Our notation is summarized in Table~\ref{tab:notation}. The units are conventional (in particular, the surface gravity is in cgs unit). For brevity we shall refer to stellar astrophysical parameters collectively as ``APs''.

\begin{table}[th]
\begin{center}
\caption{Notation. $\AG$ depends on the definition of the $G$ band and on the SED of the star, whereas $\A0$ is an extinction parameter independent of these (see section 2.2 of CBJ11). We therefore use the latter as an AP, calculating the former via an empirically-fitted relation $\AG \,=\, \A0 + y(\A0, \teff)$.}
\label{tab:notation}
\begin{tabular}{ll}
\hline
$G$  & apparent magnitude in the $G$ band (mag)\\
$\MG$ & absolute magnitude in the $G$ band (mag)\\
$\AG$ & extinction in the $G$ band (mag) \\
$\A0$ & extinction parameter (mag) \\
$\teff$  & stellar effective temperature (K) \\
$\feh$  & metallicity (dex) \\
$\logg$ & logarithm of the surface gravity (dex) \\
$\parallax$ & parallax (arcsec) \\
$q$  & $\equiv G + 5\log\parallax$ (mag) \\
$\vecp$ & a normalized SED (e.g.\ BP/RP) $\{p_i\}$, $i=1\ldots I$ \\
$\phi$ & an arbitrary astrophysical parameter\\
\hline
\end{tabular}
\end{center}
\end{table}

\section{Astrophysical parameter estimation methods}\label{sect:meth}

GSP-Phot comprises three algorithms for estimating APs which run separately from one another. All of these are described in detail in the literature, so we limit ourselves here to an overview and a description of customizations or further developments. All of the algorithms have been coded in Java.

\subsection{Support Vector Machine (SVM)}\label{sect:SVM}

The SVM is a widely-used kernel-based machine learning algorithm. 
SVMs were originally developed to perform two-class classification, but have been generalized to multi-class classification and to regression (e.g.\ \citealp{cortes95}, \citealp{burges98}). The principle of a regression SVM is to implicitly transform the data into a different data space (via a kernel) and then to fit a linear model to the data (which is typically nonlinear in the original data space).  A perfect fit is not normally possible, so the tolerance to deviations from the fit are controlled via a (hyper)parameter $\epsilon$.  This is one of three hyperparameters which must be fixed as part of the ``tuning'' process of the SVM. (The other two are the length scale in the Gaussian kernel, $\gamma$, and the regularization parameter, $C$.) We tune using a grid search. For fixed hyperparameters, the training process is a strictly convex optimization -- has a unique solution -- and results in identifying those stars, the {\em support vectors}, which lie within the tolerance range defined by $\epsilon$.  Only these are relevant in determining the APs of newly presented stars.  The training data in the present work are defined at a small number of discrete G magnitudes. SVMs perform best when trained on noisy data, and so we train a separate classifier for each discrete G.  When applying these SVM models to the test data (for which G is known), we select the model having the nearest G magnitude. 
SVM can only model one quantity at a time, so we build separate models for each AP.
We use the libSVM implementation of \cite{libsvm}.

\subsection{\ilium}

\ilium\ estimates APs via nonlinear forward modelling and iterative local search.
In an initial training phase, a multidimensional forward model is fit to a template grid to predict the flux as a function of the APs, with a separate model fit for each wavelength element (spectral band). That is, for each band we learn the function $p_i = f_i(\vecphi)$. These forward models are used in a Newton-Raphson algorithm to iteratively search the AP space in order to generate the spectrum which best matches the observed spectrum. This is a gradient-based search algorithm, in which the gradients are the sensitivities of the fluxes to the APs, $\partial p_i/\partial \phi_j$. Whereas an inverse-modelling method like SVM must implicitly learn the relevances of the spectral bands for AP estimation, this information is calculated explicitly with \ilium.

The algorithm is described in full detail in \cite{cbj10} (hereafter CBJ10), which also reported earlier results of estimating three APs from the BP/RP spectra. Essentially the same algorithm is used here with similar parameter settings, but now each forward model is a four-dimensional function over the four APs we wish to infer. As in CBJ10, the forward model is divided into {\em strong} and {\em weak} components, where $\teff$ and $\A0$ are the strong parameters and \feh\ and $\logg$ are the weak parameters (more on this later). A two-dimensional thin-plate spline is used in each case. The model is trained on the noise-free regular template grid (see section~\ref{sect:data}).
In addition to AP estimates, \ilium\ also provides covariances on these, a goodness-of-fit, and a prediction of the spectrum at the estimated APs.

\subsection{Aeneas}\label{sect:aeneas}

Aeneas is a Bayesian method which estimates the posterior probability density function (PDF) of the APs given the measured data.
In addition to using the spectrum, Aeneas also takes into account the measured parallax, $\parallax$, and apparent magnitude, $G$, of the star. These two measurements contain information on the sum of the absolute magnitude ($\MG$) and the G-band extinction ($\AG$) via the relation (continuity equation)
\begin{equation}
G + 5\log\parallax = \MG + \AG - 5 \ .
\label{eqn:ma_constraint2}
\end{equation}
Defining $q =  G + 5\log\parallax$, Aeneas infers $P(\vecphi | \vecp, q)$, where $\vecphi = (\teff, \A0, \feh, \logg)$.
The full description is given in \cite{cbj11} (hereafter CBJ11), but the basic ideas is as follows. We have three separate, but non-independent, sources of information: (1) the (normalized) spectrum, $\vecp$, which constrains $\vecphi$; (2) $q$, which constrains $\MG + \AG$; (3) the four-dimensional 
Hertzsprung--Russell diagram (HRD) -- which can be considered as the PDF $P(\MG, \teff, \feh, \logg)$ -- which constrains all of its arguments. (The normal HRD is a two-dimensional PDF $P(\MG, \teff)$.) 
The HRD we used is shown later.
After writing down an appropriate noise model for the measured quantities $\vecp$ and $q$, the
method involves marginalizing over $\MG$ to give $P(\vecphi | \vecp, q)$. 
We infer the posterior PDF using a Markov Chain Monte Carlo (MCMC) Metropolis algorithm to sample the AP space.
We take our AP estimate to be the mean of this posterior PDF.\footnote{The mean is convenient to calculate, but may limit our accuracy for low extinction stars. This is because unless the posterior PDF is infinitesimally narrow, it will extend over a finite, positive range of $\A0$, and the mean of this will always be greater than zero, even if the true $\A0$ is very close to zero.}

The advantages of this method over the previous two methods are several: (1) it provides a complete PDF over the APs, from which we can derive estimates and confidence intervals and can identify degeneracies or multimodality; 
(2) it takes into account the measurement uncertainties on all measured quantities; 
(3) it uses both the parallax and apparent magnitude self-consistently to improve the accuracy and precision of the AP estimates over what is possible with the spectrum alone (this was demonstrated in CBJ11 with Hipparcos/2MASS data); (4) it includes the prior knowledge of the HRD, such as taking into account the relative rareness of giants compared to dwarfs. This has occasionally been introduced into other AP estimation methods, but in an ad hoc way.

The PDF we derive includes the term $P(\vecp | \vecphi)$, the probability of the spectrum given the APs. We assume that each flux in each band has been measured independently and that the noise is Gaussian, in which case this term is just a product of $I$ 1-dimensional Gaussians
\begin{equation}
P(\vecp | \vecphi)  = \prod_{i=1}^{i=I} \frac{1}{\sqrt{2\pi}\sigma_{p_i}} 
                                 \exp \left( - \frac{1}{2} \left[ \frac{p_i -  f_i(\vecphi)}{\sigma_{p_i}} \right]^2 \right)
\label{eqn:likelihood}
\end{equation} 
where $f_i(\vecphi)$ is the same forward model from \ilium\ and $\sigma_{p_i}$ is the uncertainty in $p_i$. We likewise assume a Gaussian noise model for $q$ with a standard deviation depending on the estimated uncertainties in $G$ and $\parallax$ (see section 2.5 of CBJ11).

Note that we are not compelled to use $q$ or the HRD. If we leave them both out and use only the BP/RP spectrum then we are estimating $P(\vecphi | \vecp$), which we refer to as the {\em p-model}. 
Including $q$ and the HRD gives the {\em pq-model}. We provide results from both.

%

\subsection{Method comparison}

We have three algorithms which we use to produce four separate estimates of the APs (four ``methods'').  They are quite different in their approaches. SVM is trained in advance on template data to predict the APs directly as a function of a measured spectrum (an {\em inverse} model). As everything is fitted in advance, it is quick to apply. \ilium\ also involves training, but of a {\em forward} model, which is then applied iteratively in combination with the sensitivity (Jacobian) matrix to find that forward-modelled spectrum (and the APs thereof) which best resembles the measured spectrum. Aeneas determines the entire PDF over the APs given the spectrum and, optionally, the parallax, apparent magnitude and an HRD. This is naturally quite time-consuming, making this the slowest method of the three, but also the most informative. Although Aeneas uses the same forward model as \ilium\ fitted on the same training data, even the p-model generally gives better performance on account of this sampling of the PDF.

\section{Simulated data}\label{sect:data}

For the purpose of this article we use simulated data for training and testing the algorithms.  Although a lot of real stellar spectroscopic observations exist which one could attempt to use, they do not provide sufficiently dense or broad AP coverage. The required grid could only be assembled from a large number of sources, resulting in significant heterogeneity which would introduce biases, limiting the precision of AP estimation.  Even then, such a grid would not cover the full wavelength range of BP/RP.  APs must ultimately be determined from models -- and thus from synthetic spectra -- anyway, and all three AP estimation algorithms used in GSP-Phot permit us to do this directly. Synthetic spectra have the advantage that we can sample them at (or interpolate them to) arbitrary densities over the wavelength range of choice.  We must emphasize, however, that the final grids to be used in the data processing will be ``calibrated'' using a subset of stars for which accurate APs are determined via line analysis of high-resolution ground-based spectra (\citealt{heiter08}).

\subsection{Synthetic spectra}\label{sect:synspec}

\begin{figure*}
\begin{center}
\includegraphics[scale=0.7]{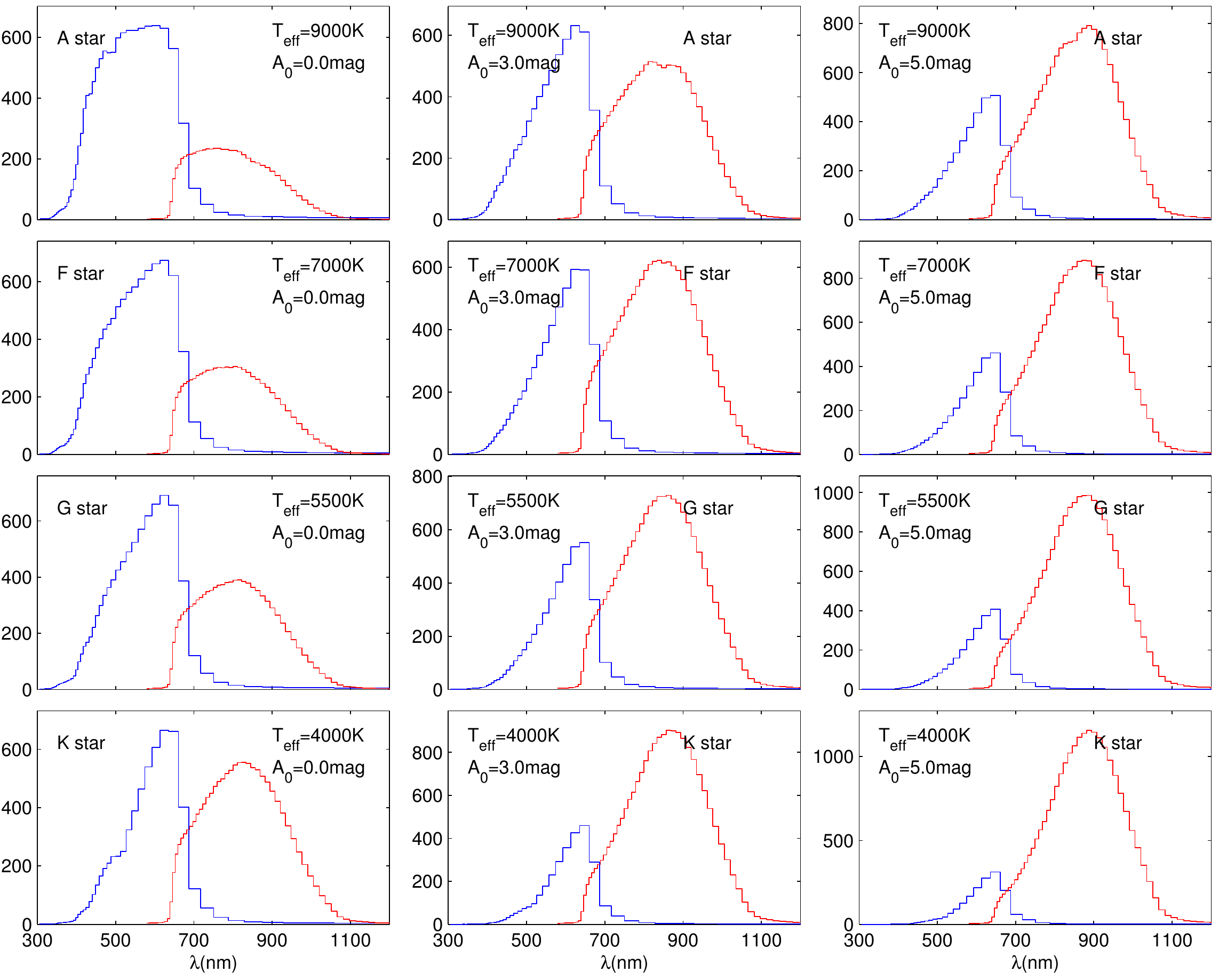}
\caption{Example (noise-free) Gaia BP/RP spectra at a common G magnitude showing how their shape depends on the strong APs $\A0$ and $\teff$. The effective temperature increases from 4000\,K to 9000\,K from bottom to top, and the interstellar extinction increases from 0\,mag to 5\,mag from left to right. The vertical axis is proportional to the number of photons collected in each band.}
\label{fig:sample1}
\end{center}
\end{figure*}

\begin{figure}
\begin{center}
\includegraphics[scale=0.7]{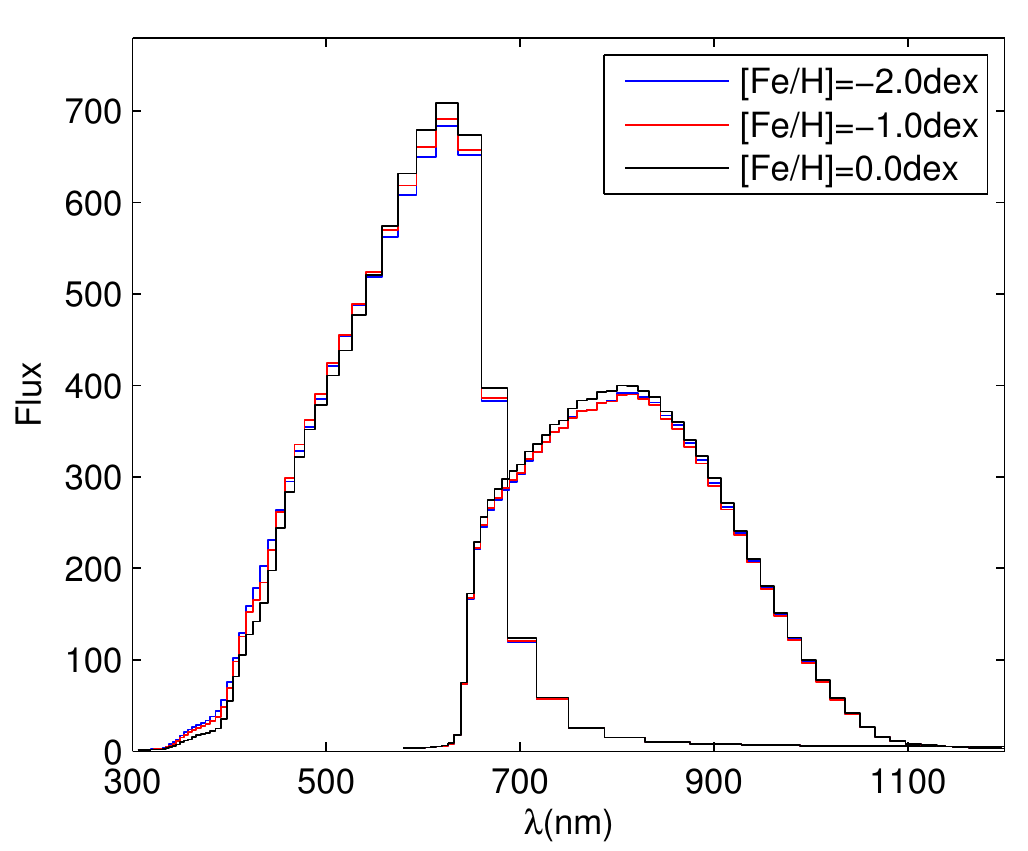}
\includegraphics[scale=0.7]{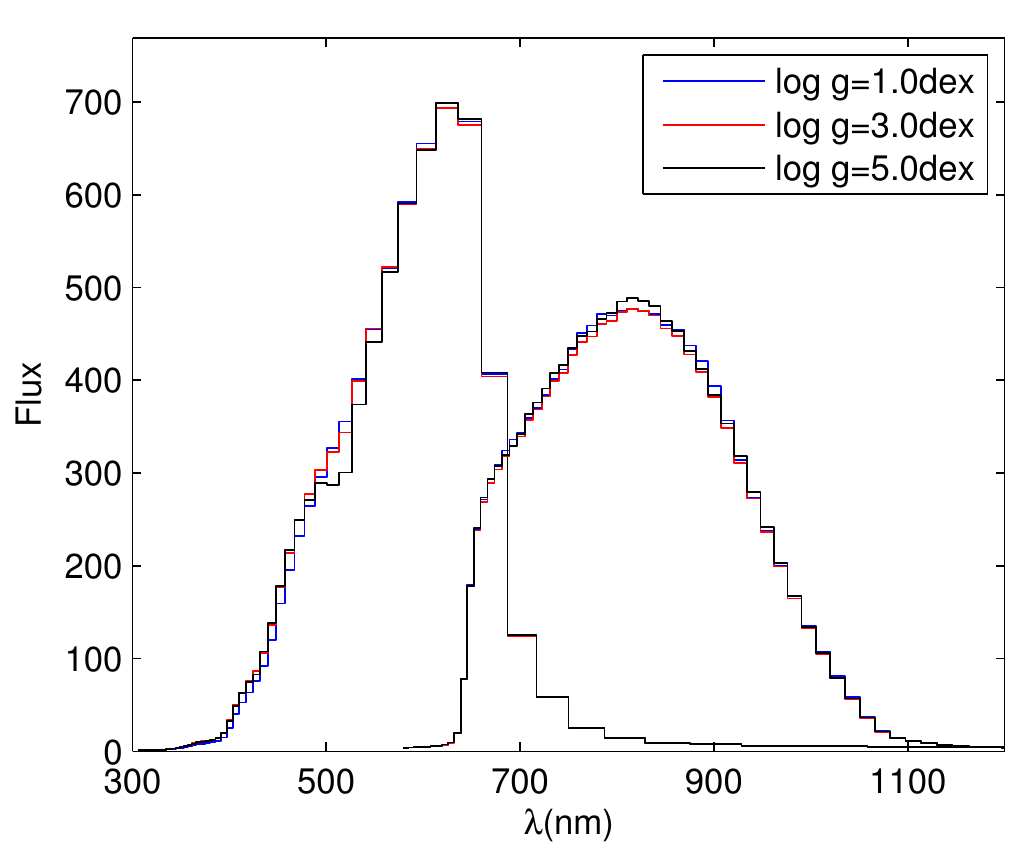}
\caption{Example (noise-free) BP/RP spectra showing how their shape depends on the weak APs $\logg$ and \feh\ at $\teff$\,=\,5000\,K and $\A0$\,=\,0\,mag. The upper panel shows spectra with various \feh\ at fixed $\logg$\,=\,4\,dex, the lower panel spectra for various $\logg$ at fixed \feh\,=\,0\,dex The vertical axis is proportional to the number of photons collected in each band.}
\label{fig:sample2}
\end{center}
\end{figure}

The input stellar library is a version of the Phoenix library (\citealp{brott05}). These we redden to a range of interstellar extinctions (using the law of \citealt{cardelli89}) and then convert into Gaia BP/RP spectra using the GOG (Gaia Object Generator) instrument simulator (\citealt{luri05}, \citealt{isasi09}).  The resulting libraries are described in \cite{sordo12}. (These are the  DPAC cycle 8 simulations.) Due to the broad line spread function of the instrument, the resulting spectra have very low resolution, in which hardly any absorption lines are visible, and the continuum is strongly modulated by the instrument response.  The spectra are recorded separately in the two channels BP and RP (each an array of CCD detectors) with 60 data samples in each. We  refer to these samples as ``bands'', as they are like (overlapping) medium band filters.
BP covers the wavelength range 330--680\,nm with a resolution (=$\lambda / \Delta\lambda$) varying from 85 to 13 across this range. RP covers 640--1050\,nm with resolution from 26 to 17 \citep{debruijne12}. ($\Delta\lambda$ is defined as the 76\% energy width of the line spread function.)

Figure~\ref{fig:sample1} shows how the spectra depend on $\teff$ and $\A0$. Note how the flux in the blue channel (BP) is strongly suppressed at higher extinctions. As is well known, the effects of temperature and extinction are hard to disentangle at low resolution. For instance, it is hard to spot the difference between the spectrum with $\teff=4000$\,K and $\A0=3$\,mag (bottom middle panel) and the one with $\teff=5500$\,K and $\A0=5$\,mag.  The difficulties this degeneracy presents have been discussed in CBJ10 using \ilium, and will be taken up again in section~\ref{subsect:a0deg}.

The variance in the spectra due to $\teff$ and $\A0$ dominates over that of \feh\ and $\logg$, as can be seen in Figure~\ref{fig:sample2}, and for this reason we refer to former pair as {\em strong} APs and the latter pair as {\em weak} APs. On account of this we expect to be able to estimate the strong APs to a higher accuracy in the presence of noise.
For faint stars (lower SNRs), the variation due to $\feh$ or $\logg$ may be flooded by noise, making these APs very difficult to estimate. This will be quantified in this work. (For more plots of the BP/RP spectra, see Figures 6--9 of CBJ10.)

\subsection{Noise models}\label{sect:noise}

\begin{figure}
\begin{center}
    \includegraphics[width=0.40\textwidth]{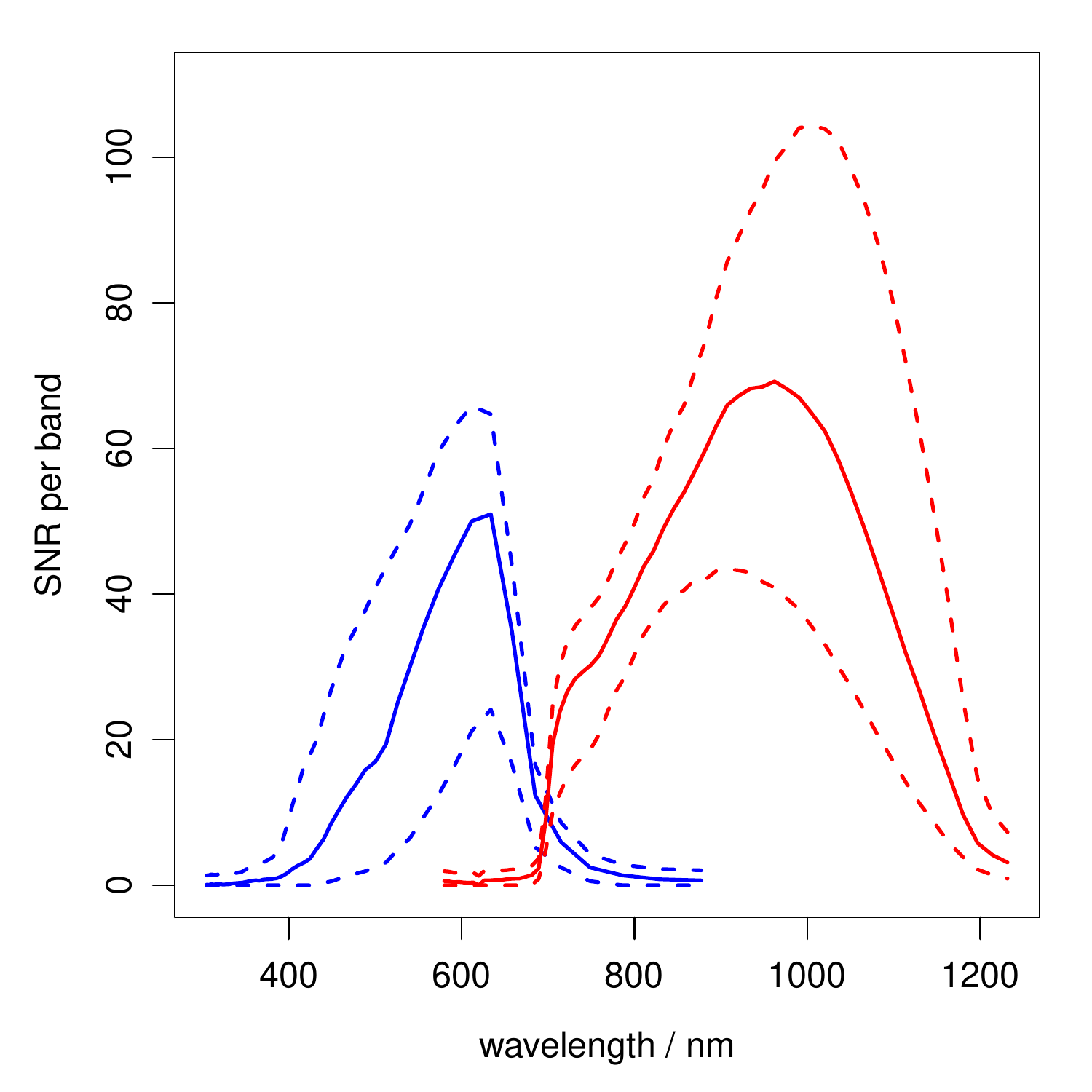} 
    \caption{The median signal-to-noise ratio (SNR) per band (solid line) and the 0.1 and 0.9 quantiles
      (dashed lines) across the sample of 2000 stars in the G=19 test set. The SNR at G=15 is 10--25 times larger (depending on wavelength).}
\label{fig:snr_G19}
\end{center}
\end{figure}

All tests in this paper are done on noisy test data.  Given the way Gaia scans the sky, all sources receive (to zeroth order) the same amount of integration time over the course of the five year mission, so the SNR depends primarily on the G magnitude. (The integration time depends on the number of observation transits, which is a function of sky position and varies by a factor of a few. We adopt here a nominal 72 transits per source.) The spectrophotometric noise model in GOG
takes into account the Poisson noise in the source and background, as well as the detector (CCD) noise and some additional photometric processing noise \citep{jordi09}. A 20\% error margin is added to accommodate unmodelled random errors, in particular imperfect geometric calibration and charge transfer inefficiency \citep{sartoretti11}. Systematic errors are currently neglected.
The range of typical SNRs per band for the test set at G=19 (described in the next subsection) is shown in Fig.~\ref{fig:snr_G19}. Note that this set has a large fraction of high extinction stars. (See also Figure 10 of CBJ10, which is for a sample of zero extinction stars).
The actual SNR which Gaia will deliver as a function of magnitude may well differ from what is assumed here. In that case the results at a specified magnitude would apply to a slightly higher or lower magnitude.

We use a parallax noise model to generate noisy parallaxes.  The standard deviation of the parallax ranges from around 10--25\,\uas\ (depending on the colour of the star; it is better for redder stars) at G=15 to 100--330\,\uas\ at G=20.  For the G=15 test set, the median SNR in the parallax, $\langle \varpi / \sigma(\varpi) \rangle$, is 900, and for the G=19 test set it is 12.  (The former value is large because the test set contains many nearby stars, which in turn is due to the large extinctions.)

\subsection{Train/test data sets}\label{sect:grids}

\begin{figure}
\begin{center}
\includegraphics[scale=0.7]{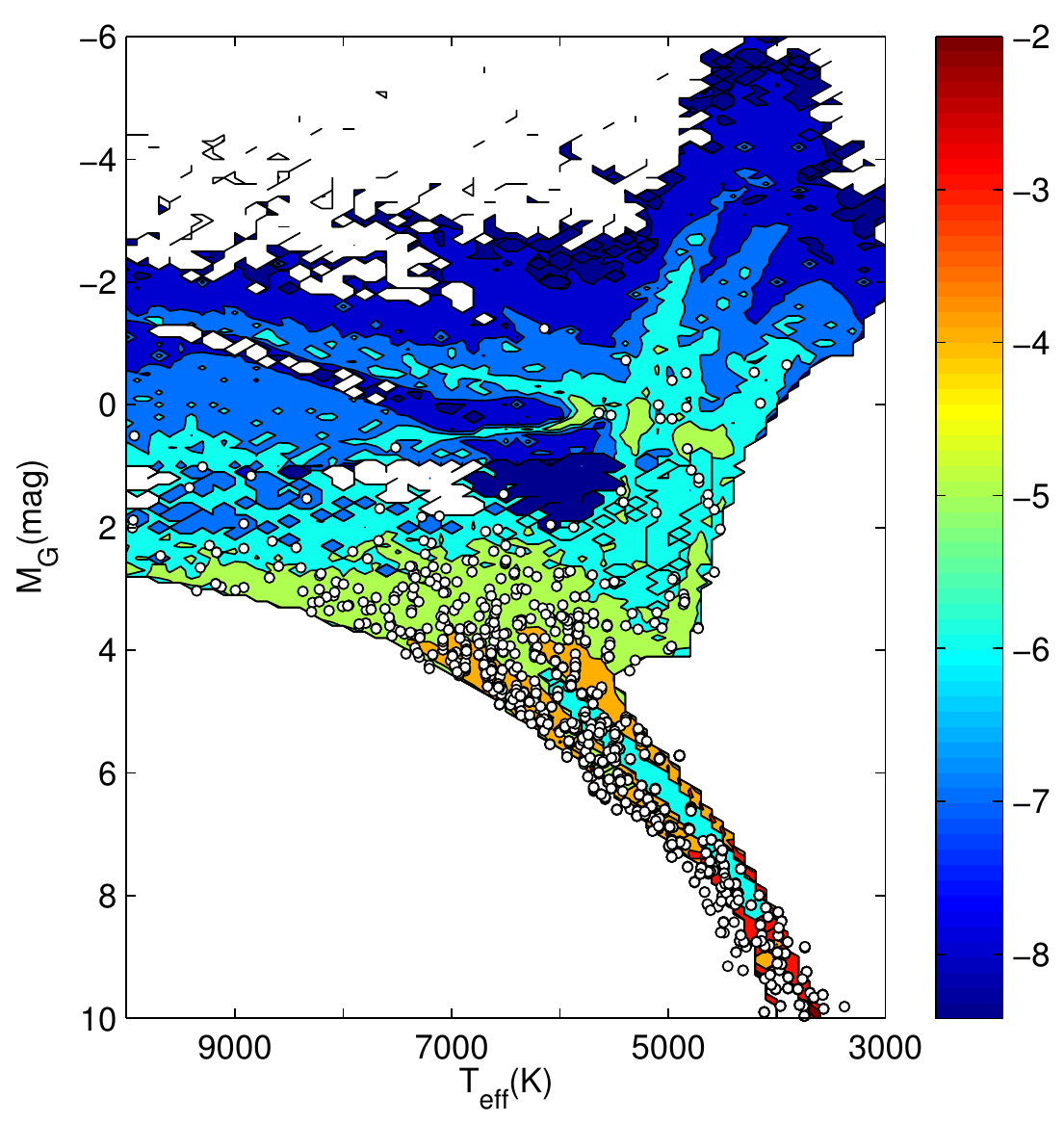}
\caption{The colour map shows the HRD (as a normalized probability density function on a log scale) used as part of the procedure for selecting stars to build the test data sets. This HRD is also used as the prior in Aeneas. The white circles identify the 2000 stars in test data set at G=15. (The APs in the other random grid data sets, at G=19 and at mixed magnitudes, are statistically the same.)}
\label{fig:hrd}
\end{center}
\end{figure}

\begin{figure}
\begin{center}
\includegraphics[scale=0.7]{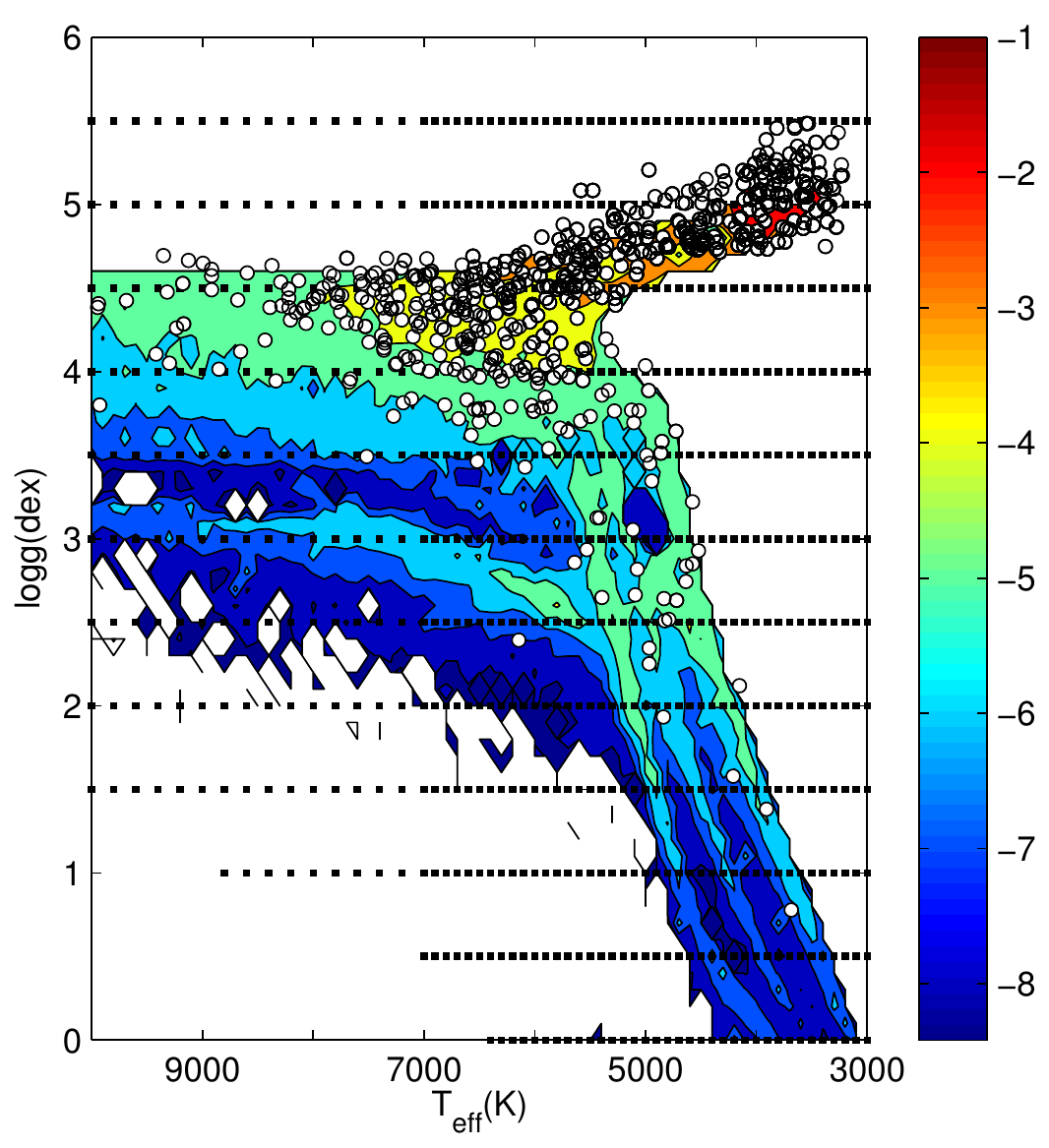}
\caption{The HRD and the white circular points are as Fig.~\ref{fig:hrd}, but now in $\teff$--$\logg$ space. The black rectangular points in rows of constant $\logg$ are the values in the regular grid.}\label{fig:teffloggdist}
\end{center}
\end{figure}

\begin{figure}
\begin{center}
\includegraphics[scale=0.8]{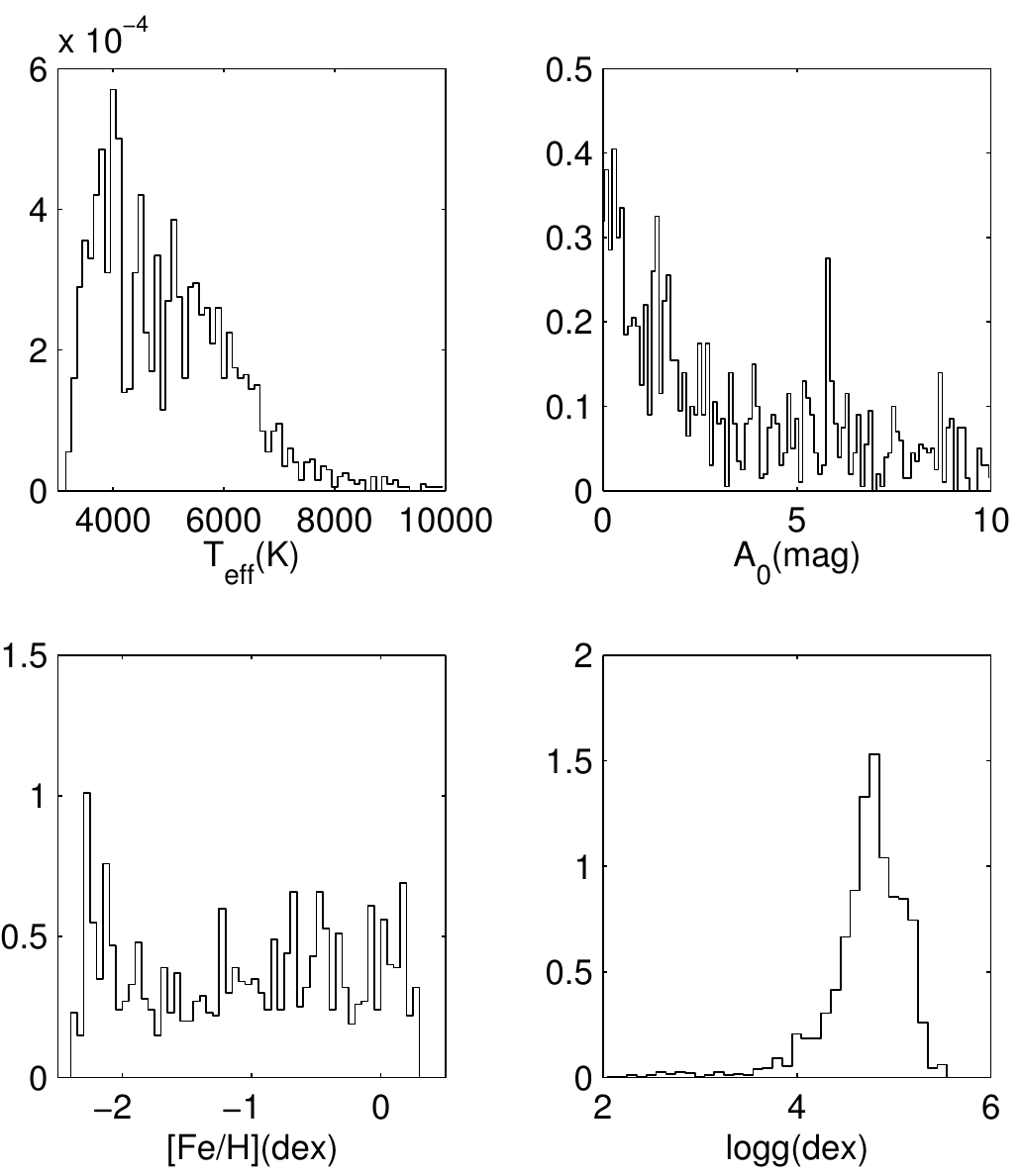}
\end{center}
\caption{The (normalized) distributions over the APs in the test data set at G=15. (The APs in the other random grid data sets, at G=19 and at mixed magnitudes, are statistically the same.)}\label{fig:apdist}
\end{figure}

The synthetic spectra cover the range $\teff$\,=\,3000--10\,000\,K, $\A0$\,=\,0--10\,mag, \feh\,=\,-2.5 to +0.5\,dex, and $\logg$\,=\,-0.5 to +5.5\,dex. We use them to construct two distinct grids:
\begin{itemize}

\item A {\em regular grid} -- one with the APs at regularly spaced intervals -- in which all BP/RP spectra are noise-free.  (They are all scaled to G=15, an arbitrary choice.)  The distribution in $\teff$ and $\logg$ is shown in Fig.~\ref{fig:teffloggdist} (black rectangular points).  The $\A0$ values are 0, 0.1, 0.5, 1, 2, 3, 4, 5, 8, 10\,mag, $\logg$ ranges from -0.5 to +5.5\,dex in steps of 0.5\,dex, and \feh\ from -2.5 to +0.5\,dex in steps of 0.5\,dex. This grid is used for building the forward models in \ilium\ and Aeneas. (The $\teff$, $\logg$, and \feh\ values are just those at which the Phoenix spectra were calculated.)  

\item {\em Random grids} with APs at arbitrary values. The input spectra for an initial random grid are generated by locally interpolating the input spectra in the regular grid (\citealt{vallenari08}).  (This interpolation is done on the high resolution spectra, i.e.\ prior to simulation in GOG.)  This grid, like the regular one, covers a wide range of APs, but includes some non-physical combinations of APs, i.e.\ combinations corresponding to stars which do not really exist. It also over-represents rare stars.  We therefore select stars from this initial random grid based on sampling from a Hertzsprung--Russell diagram (HRD). (This is possible because the simulated sources have been assigned absolute magnitudes; see \citealt{vallenari10} and \citealt{sordo12}).  For this we use the HRD from CBJ11, shown in Fig.~\ref{fig:hrd} (colour scale). This HRD reflects the stellar population one would expect after a long period of star formation with a typical initial mass function, so both the HRD and the selection are highly non-uniform (the white points in the figure; the discrepancy will be discussed below). As directly sampling from the HRD produces relatively few giants and hot stars (which would lead to poorer SVM training and poorer test statistics), we actually modify the selection somewhat in order to augment these types of objects.
The result of this selection we call a {\em random grid}, and such grids, which all have the same AP distribution statistically speaking, are used to test all methods and to train the SVM. 

\end{itemize}
AP estimation accuracy depends strongly on the SNR. To investigate this we construct three test data sets from the random grids
\begin{enumerate}
\item 2000 stars at G=15;
\item 2000 stars at G=19;
\item 2000 stars assigned random magnitudes drawn uniformly between G=6 and G=20, which we will call the {\em mixed magnitude} data set.
\end{enumerate}
The performance of the methods is evaluated using these three data sets. 
Note that the extinction distribution in these test data sets is not representative of any astrophysically obvious sample which Gaia will observe. A broad distribution has been adopted for all APs in order to explore how AP accuracy varies with the APs.

We also use random grid data sets for training the SVM.  As mentioned in section~\ref{sect:SVM}, SVM performs best when the noise in the training set is matched to the noise in the test set.  Thus for processing the test set at G=15, we use an SVM which has been trained on a set of 10\,000 stars taken from the random grid, also at G=15. We do the same for G=19. For processing the mixed magnitude test set, we use a set of SVM models, each trained on 10\,000 stars at G magnitudes of 9, 15, 17, 18, 18.5, 19, 19.5, and 20. Each star in the mixed magnitude test data set is sent to the SVM model with the nearest G magnitude.  This kind of G matching approach will be adopted for the real mission data processing.

Note that the SVM is trained and tested on data with similar AP distributions (but not the same stars). While this is expected to optimize SVM performance, it makes the SVM somewhat sensitive to the training data distribution. That is, if the SVM is then used on a data set with a very different AP distribution (e.g.\ one with mostly giants, which are relatively rare in the random grids), it may not perform so well. In contrast, \ilium\ and Aeneas, being based on forward modelling, are less sensitive to this training data distribution issue.

Each BP/RP spectrum is individually normalized (by dividing the spectrum by its integral) before being used by any algorithm for training or testing. This removes the overall flux level.

Figure~\ref{fig:apdist} shows the distributions over all APs in the test set at G=15.
Figure~\ref{fig:teffloggdist} shows its distribution in the $\teff$--$\logg$ space, along with the HRD.
We see that the distribution of the test data is not entirely consistent with the HRD, there being an apparent misalignment by a few tenths of a dex in $\logg$.
There are three reasons for this discrepancy: (1) for technical reasons we could not draw arbitrary stars directly from the HRD; for each draw we had to select the simulated spectrum with the nearest $(\teff, \MG)$ from a pre-generated initial random grid; (2) the $\MG$ of the stars in this pre-generated initial random grid are not exact (as they are assigned by matching to isochrones); (3) different model atmospheres were used to build the spectral library (Phoenix) and to construct the HRD. 
Consequently the parallaxes may not be entirely consistent with the APs.
This inconsistency will not affect the results from SVM, \ilium\ or Aeneas p-model, as they only use the spectra: The spectra and APs in these training and test data sets are self-consistent within the limits of the interpolation used to construct the initial random grids. However, this inconsistency may bias the Aeneas pq-model, because it uses both the parallax/HRD and the spectrum to infer the APs, relying on the self-consistency implicit in equation~\ref{eqn:ma_constraint2}. 
This needs to be improved in future work, ideally by building the HRD and test data sets from a common library.\footnote{Ideally we would generate spectra directly from a combined stellar evolution/atmosphere model for given mass, metallicity and age, and then derive $\teff$, $\logg$ and \feh, as well as $\MG$. Placing the star at a given distance (parallax), and applying extinction, we can then derive $G$. In practice, however, we have pre-defined spectral libraries, and for performance estimation we want to generate stars at constant $G$. This makes certain approximations unavoidable.}

\section{Results (predicted performance)}\label{sect:res1}

\begin{table}
\caption{Performance (mean absolute residual) for stars at $G=15$\,mag with true $\A0<1$\,mag.}\label{tab:resultsg15}
\begin{tabular}{cccccc}
\hline
AP & Sample & SVM & \ilium\ & Aeneas & Aeneas \\
&&&& (p-model)&(pq-model)\\
\hline
& All stars & 59 & 110 & 72 & 71 \\
& A stars & 132 & 111 & 143 & 139 \\
$\epsilon_{\teff}$/K& F stars & 66 & 84 & 82 & 78 \\
& G stars & 57 & 87 & 67 & 69 \\
& K stars & 53 & 152 & 70 & 69 \\
& M stars & 50 & 70 & 56 & 53 \\
\hline
& All stars & 0.04 & 0.09 & 0.07 & 0.07 \\
& A stars & 0.03 & 0.04 & 0.04 & 0.04 \\
$\epsilon_{\A0}$/mag& F stars & 0.03 & 0.04 & 0.04 & 0.04 \\
& G stars & 0.03 & 0.05 & 0.05 & 0.05 \\
& K stars & 0.04 & 0.16 & 0.09 & 0.09 \\
& M stars & 0.07 & 0.07 & 0.09 & 0.09 \\
\hline
& All stars & 0.20 & 0.16 & 0.15 & 0.14 \\
& A stars & 0.68 & 0.34 & 0.48 & 0.40 \\
$\epsilon_{\feh}$/dex& F stars & 0.23 & 0.12 & 0.11 & 0.09 \\
& G stars & 0.14 & 0.10 & 0.09 & 0.09 \\
& K stars & 0.14 & 0.21 & 0.14 & 0.12 \\
& M stars & 0.28 & 0.13 & 0.23 & 0.23 \\
\hline
& All stars & 0.17 & 0.34 & 0.28 & 0.20 \\
& A stars & 0.13 & 0.19 & 0.19 & 0.19 \\
$\epsilon_{\logg}$/dex& F stars & 0.18 & 0.29 & 0.24 & 0.21 \\
& G stars & 0.19 & 0.31 & 0.23 & 0.16 \\
& K stars & 0.16 & 0.41 & 0.31 & 0.20 \\
& M stars & 0.15 & 0.30 & 0.36 & 0.25 \\
\hline\end{tabular}\end{table}

\begin{table}
\caption{Performance (mean absolute residual) for stars at $G=19$\,mag with true $\A0<1$\,mag.}\label{tab:resultsg19}
\begin{tabular}{cccccc}
\hline
AP & Sample & SVM & \ilium\ & Aeneas & Aeneas \\
&&&& (p-model)&(pq-model)\\
\hline
& All stars & 199 & 451 & 209 & 294 \\
& A stars & 590 & 648 & 628 & 933 \\
$\epsilon_{\teff}$/K& F stars & 249 & 503 & 261 & 416 \\
& G stars & 226 & 502 & 233 & 337 \\
& K stars & 185 & 523 & 197 & 251 \\
& M stars & 97 & 148 & 99 & 138 \\
\hline
& All stars & 0.16 & 0.30 & 0.16 & 0.23 \\
& A stars & 0.13 & 0.19 & 0.16 & 0.30 \\
$\epsilon_{\A0}$/mag& F stars & 0.13 & 0.26 & 0.14 & 0.21 \\
& G stars & 0.14 & 0.28 & 0.14 & 0.21 \\
& K stars & 0.20 & 0.37 & 0.20 & 0.26 \\
& M stars & 0.13 & 0.22 & 0.14 & 0.19 \\
\hline
& All stars & 0.37 & 0.68 & 0.34 & 0.47 \\
& A stars & 0.72 & 0.99 & 0.69 & 0.83 \\
$\epsilon_{\feh}$/dex& F stars & 0.37 & 0.62 & 0.30 & 0.42 \\
& G stars & 0.30 & 0.69 & 0.34 & 0.46 \\
& K stars & 0.37 & 0.81 & 0.35 & 0.47 \\
& M stars & 0.38 & 0.39 & 0.31 & 0.44 \\
\hline
& All stars & 0.19 & 1.37 & 0.56 & 0.32 \\
& A stars & 0.19 & 0.87 & 0.91 & 0.88 \\
$\epsilon_{\logg}$/dex& F stars & 0.24 & 1.35 & 0.54 & 0.39 \\
& G stars & 0.22 & 2.08 & 0.61 & 0.40 \\
& K stars & 0.17 & 1.40 & 0.59 & 0.24 \\
& M stars & 0.13 & 0.57 & 0.43 & 0.28 \\
\hline
\end{tabular}
\end{table}

\subsection{Overview of the results}\label{subsec:overview}

\begin{figure*}
\begin{center}
\includegraphics[scale=0.8]{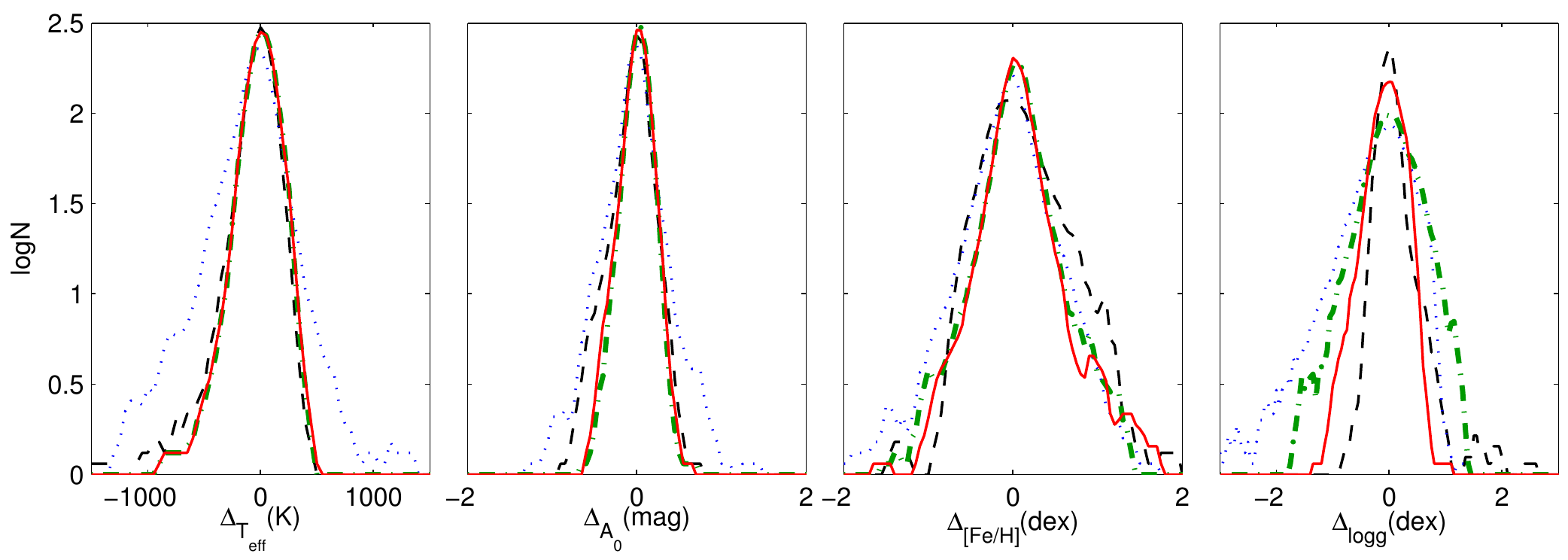}
\caption{The logarithmic histogram of the residuals of $\teff$, $\A0$, $\feh$, and $\logg$ for all stars with $G=15$\,mag.
The colours indicate the method: SVM (black/dashed), \ilium\ (blue/dotted), Aeneas p-model (green/dashed--dotted), Aeneas pq-model (red/solid).}
\label{fig:histres}
\end{center}
\end{figure*}

The main objective is to achieve accurate AP estimation, so in analysing performance we focus on the residuals, predicted AP minus true AP.  As accuracy is a strong function of magnitude and the true AP values, it can be misleading to summarize the performance with just a few numbers, as any average depends strongly on the distribution of APs in the test data set. We therefore show various plots and tables for various AP ranges. We also provide star-by-star results in a set of online tables (see appendix).

Figure~\ref{fig:histres} permits a quick comparison of the methods by showing the distributions of the residuals for the G=15 test set.  
Table~\ref{tab:resultsg15} summarizes the accuracy of each AP for low extinction stars ($\A0<1$\,mag) for different ranges of the true $\teff$\ (``spectral types'') for each of the four methods, at G=15.  The spectral types are defined as the following $\teff / K$ ranges (with the number of stars of each type in the full test set / in the $\A0<1$ subset given in parentheses): 7500--10000 A stars (68/22); 6000--7500 F stars (336/106); 5250--6000 G stars (361/113); 3750--5250 K stars (932/218); 3000--3750 M stars (303/96).  Defining the residual for AP $\phi$ as $\Delta_\phi = \phi_{\rm est} - \phi_{\rm true}$, the first of our statistics for summarizing performance over a set of stars is the mean absolute residual (MAR), $\epsilon_\phi = \langle | \Delta_\phi | \rangle$.  Similar metrics are possible, such as the root-mean-square (RMS), but the MAR is less sensitive to outliers, so is more representative of the bulk of the residual distribution. Table~\ref{tab:resultsg19} summarizes the accuracy at G=19.  The dependence of the performance on G will be discussed in section~\ref{sect:disc}.

The MAR is a measure of the total error. It is useful to know how much of this is a systematic error. We characterize that using our second statistic, the mean residual (MR), $\langle \Delta_\phi \rangle$.

The reader can draw his/her own conclusions from these tables, but we draw attention to a few features.

First, SVM appears to have the best performance for almost all APs given $\A0<1$\,mag. 
We will see that this is not the case for all APs at higher extinctions, however.
$\logg$ also appears to be estimated best by SVM. However, Figure~\ref{fig:histres} shows that the residuals from SVM have a positive skew, i.e.\ a systematic over-estimation.

Second, $\feh$ can be estimated more accurately for cooler stars than for hotter stars with all algorithms. This well-known result is a consequence of metal lines being less prominent in the stellar spectra of hotter stars. Even though we have no explicit lines in the BP/RP spectra, metallicity is still exerting an effect on the pseudo continuum.

Third, at G=15 Aeneas pq-model is slightly better than p-model for $\logg$, while for the other three APs the two models achieve similar performance. At G=19 pq-model gives rise to considerably better accuracy than p-model in 
$\logg$ for FGKM stars, but is considerably worse in the other APs. This will be explained in section~\ref{subsect:ppqcompare}.

In the following sections we look in more detail into the AP estimation accuracy and its dependence on the true APs.

\subsection{$\teff$ estimation}\label{subsect:teff}

\begin{figure*}
\begin{center}
\includegraphics[width=15cm]{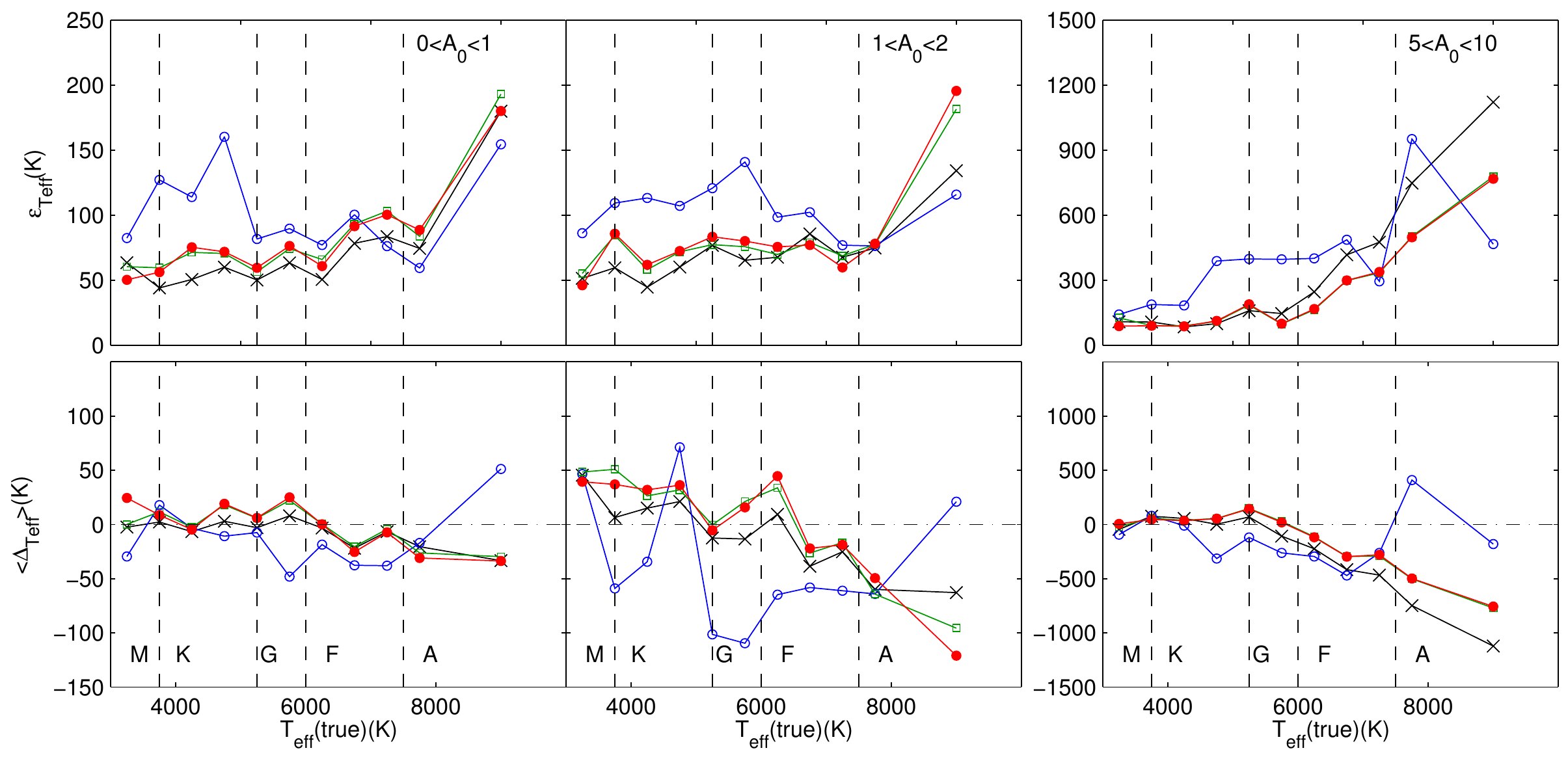}
\caption{The errors in estimated $\teff$ as a function of true $\teff$ for stars with G=15\,mag. The top row shows the mean absolute residual, $\epsilon_{\teff}$. The bottom row shows the mean residual, $\langle \Delta_{\teff} \rangle$. The columns from left to right are for stars with $\A0<1$, $1<\A0<2$, and $5<\A0<10$, respectively. 
The colours/symbols indicate the method: SVM (black crosses), \ilium\ (blue open circles), Aeneas p-model (green rectangles), Aeneas pq-model (red closed circles).
The dashed vertical lines delineate the spectral types A, F, G, K, M. Note the change in scale for the highest extinction panels.}
\label{fig:teff}
\end{center}
\end{figure*}


Figure~\ref{fig:teff} shows the dependence of the $\teff$ MAR (upper row) and MR (lower row) on the true $\teff$ for each of the four methods at G=15, for three different ranges of the true extinction. 
For most of the parameter range, SVM and Aeneas show similar performance, with \ilium\ being somewhat less accurate. The error in $\teff$ increases significantly in the final temperature bin (A stars).  
Such a phenomenon is sometimes seen when a training grid is much sparser in that part of the parameter space. Although this is the case here (see Fig.~\ref{fig:teffloggdist}), we believe this is only a weak effect here.
It may be in part a consequence of having used $\log(\teff)$ in the models rather than $\teff$: a constant error in the former corresponds to an increasing error in the latter as $\teff$ increases. (So the increase we see here is actually more modest in $\log(\teff)$.)  But probably most of the trend is real. Nonetheless, at the lower extinctions the algorithms are still able to estimate $\teff$ to an accuracy of 1--2\%.

The mean residual in SVM and Aeneas also show a correlation with $\teff$, the temperature of A/F stars being underestimated by both methods at higher extinctions. For the A stars with $\A0<1$\,mag the underestimation for the algorithms is less than 50\,K, while for the A stars with $5<\A0<10$ it grows to 500--1000\,K.
Given that this systematic error is actually a significant part of the total error in the two higher extinction bins, we might think that we could correct for this systematic error. Unfortunately this is rarely possible, for the following reason. These plots show the residuals as a function of the {\em true} $\teff$, which is unknown in a real application. Plotting instead the residuals as a function of the {\em inferred} $\teff$, there is no significant signal which we can use to apply a useful correction (see appendix A of \cite{cbj09-043} for an example).

We see that the Aeneas p-model (green lines) and pq-model (red lines) show very similar performance on $\teff$, indicating that the parallax and the HRD prior used in pq-model do not improve the performance for $\teff$ at G=15 (at least, not when averaged over the AP ranges used to make this plot).  But we will see below that it does improve the $\logg$ accuracy.

\subsection{$\A0$ estimation and the $\teff$--$\A0$ degeneracy}\label{subsect:a0deg}
\begin{figure*}
\begin{center}
\includegraphics[width=15cm]{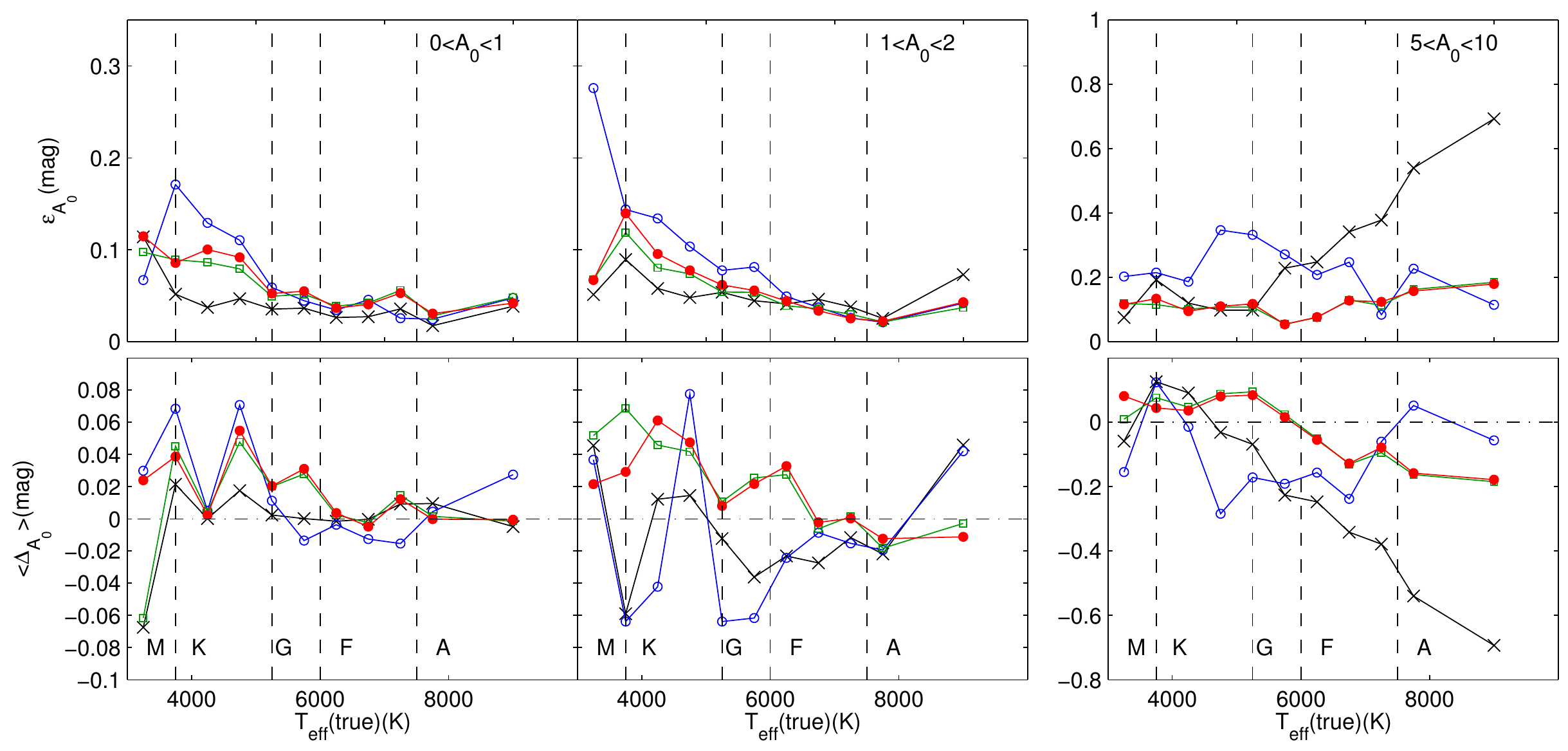}
\caption{The errors in estimated $\A0$ as a function of true $\teff$ for stars with G=15\,mag. The top row shows the mean absolute residual, $\epsilon_{\A0}$. The bottom row shows the mean residual, $\langle \Delta_{\A0} \rangle$. The columns from left to right are for stars with $\A0<1$, $1<\A0<2$, and $5<\A0<10$, respectively. 
The colours/symbols indicate the method: SVM (black crosses), \ilium\ (blue open circles), Aeneas p-model (green rectangles), Aeneas pq-model (red closed circles).
The dashed vertical lines delineate the spectral types A, F, G, K, M. Note the change in scale for the highest extinction panels.}
\label{fig:a0}
\end{center}
\end{figure*}

Figure~\ref{fig:a0} shows the performance of $\A0$.  It is encouraging for the Gaia science case to see that $\A0$ can be estimated to better than 0.2\,mag for A--M stars with extinctions even up to 10\,mag, and that for the lower extinctions an accuracy of better than 0.05\,mag is possible for most of the $\teff$ range. This will of course only hold for the real Gaia data if the extinction law we have used is true to this level of accuracy.

For all algorithms the MAR is significantly larger for K/M stars when the extinction is low. 
No single algorithm turns out to be the best. SVM is somewhat better than the other methods for G/K stars for $\A0<2$\,mag, but is much poorer on the A/F stars at $5<\A0<10$\,mag, in terms of both the systematic (MR) and total (MAR) errors, with the error increasing with increasing $\teff$. We saw a similar correlation in the residuals
for $\teff$ (Figure~\ref{fig:teff}). This implies that there is degeneracy between $\teff$ and $\A0$ which the SVM is much less able to accommodate than the other algorithms. 
We already observed in Figure~\ref{fig:sample1} the presence of this degeneracy in the spectra. This is intrinsic to the data: given the lack of resolved lines sensitive only to $\teff$, it will not be possible to remove this with any algorithm.
SVM may be more affected on account of it trying to solve an inverse mapping problem, from the data to the APs. This is not necessarily single-valued (even with the kernel function), in which case SVM cannot fit the correct function. \ilium\ and Aeneas, in contrast, rely on forward modelling (which is single-valued), which is essentially used to suggest candidate solutions (model spectra/APs). These models are also affected by the degeneracy, but at least their candidate solutions are plausible (if the forward modelling is accurate enough), and this may work to reduce the {\em bias} in their solutions. Aeneas at least has the advantage that by calculating a PDF, it can characterize the degeneracy, and by permitting the HRD/parallax to be used, can also reduced it (see CBJ11).

\begin{figure}
\begin{center}
\includegraphics[scale=0.7]{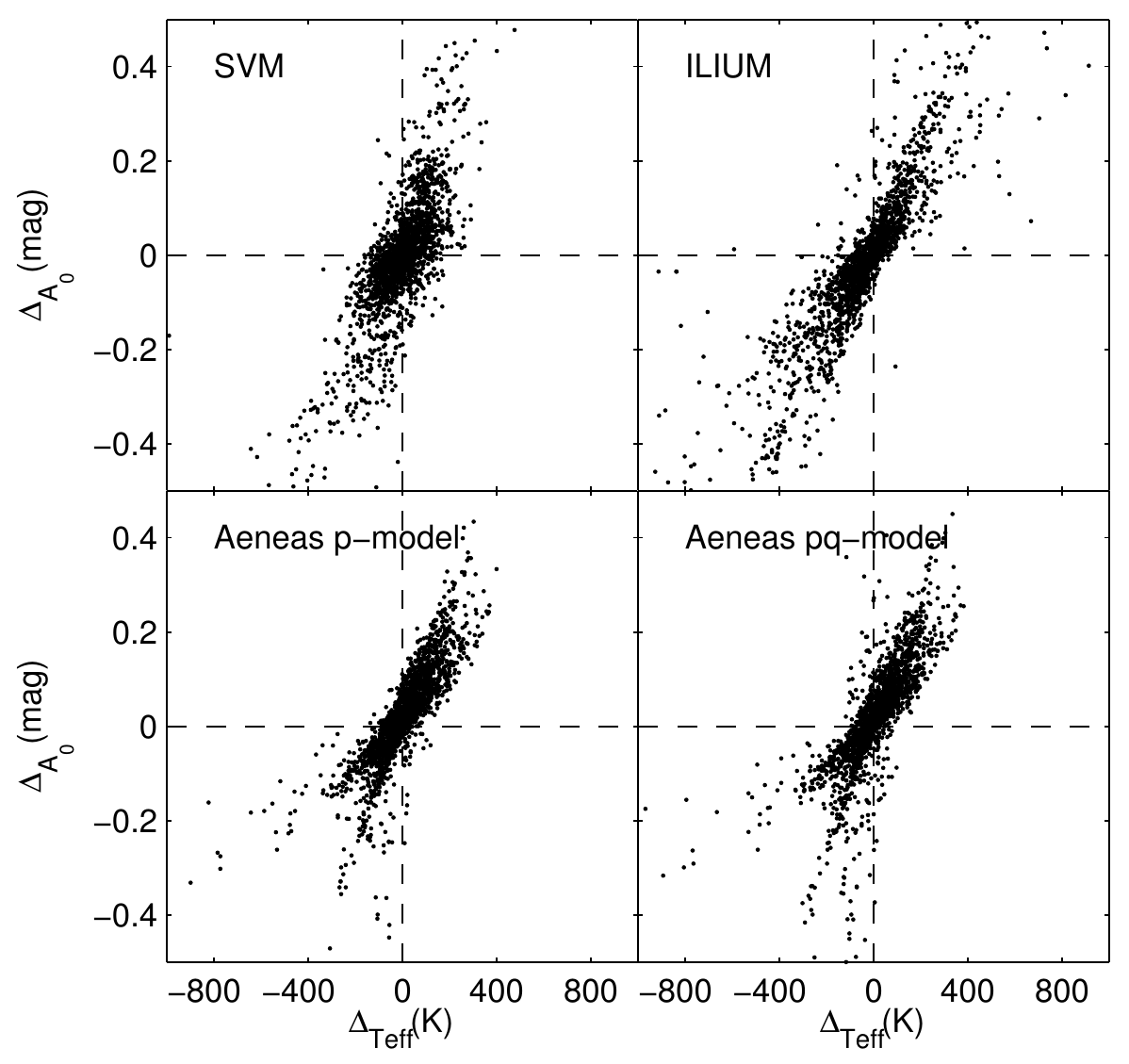} 
\includegraphics[scale=0.7]{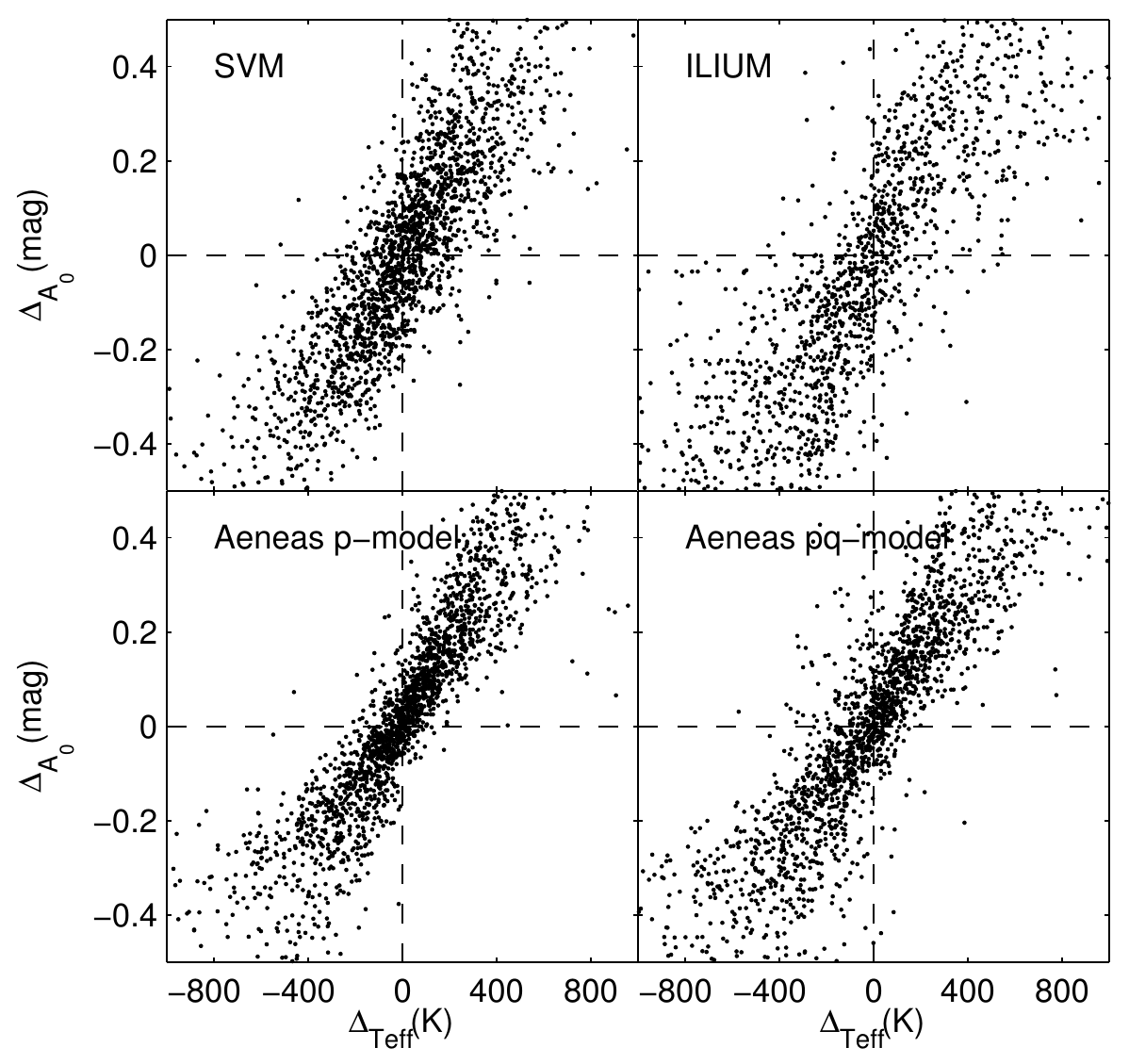} 
\caption{The correlation between the $\teff$ and $\A0$ residuals for stars with $G=15$\,mag (upper panels) and $G=19$\,mag (lower panels) for all four methods.}\label{fig:degteffa0}
\end{center}
\end{figure}

Figure~\ref{fig:degteffa0} shows the correlation between the residuals for all four methods at both G=15 and G=19. The degeneracy is indeed present for all algorithms, and is more severe at fainter magnitudes.
However, the spread in the residuals is larger for SVM than for \ilium\ or Aeneas. 

\begin{figure}
\begin{center}
\includegraphics[scale=0.7]{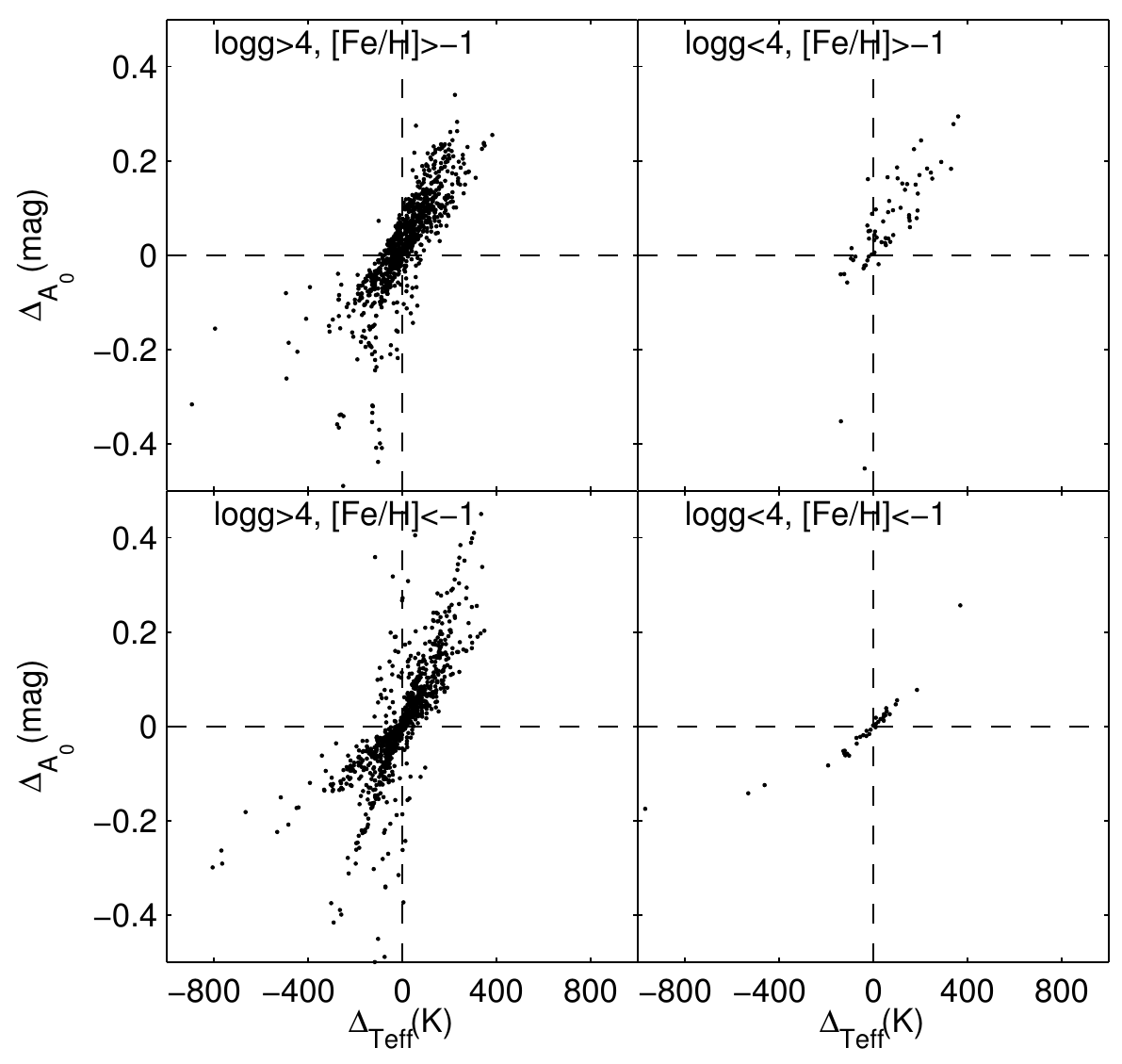}
\caption{The correlation between the $\teff$ and $\A0$ residuals from Aeneas pq-model for stars with $G=15$\,mag for
the four combinations of metal-poor/metal-rich and dwarf/giant (the four panels).}\label{fig:degteffa0-2}
\end{center}
\end{figure}

Figure~\ref{fig:degteffa0-2} shows the same correlation plot (for Aeneas pq-model) but now separately for the four combinations of metal-poor/metal-rich and dwarf/giant. We see that for metal rich giants the $\teff$--$\A0$ degeneracy is biased to only {\em overestimating} these APs. This may be related to the temperature sensitive Mgb+MgH molecular feature at around 520\,nm, which is strong in cool dwarfs (the ``dent'' in the BP spectrum in the bottom left panel of Figure~\ref{fig:sample1})
but absent in hotter stars and in giants. The cooler the star, the stronger (deeper) the feature. Assuming the algorithms are making use of this feature, then when the dip is absent -- as it is in all giant stars -- in an otherwise cool star spectrum, the algorithm incorrectly interprets this as an indication of a reddened, hotter star spectrum, thus overestimating $\teff$ and $\A0$ for (metal rich) giant stars.
As the feature is also weak or even absent in metal poor stars, we might expect a systematic overestimation of $\teff$ for these too. We do not see this, however (lower two panels of Figure~\ref{fig:degteffa0-2}), presumably because it is absent
for all $\teff$, so $\teff$ estimation is not making use of (the presence or absence of) this feature at all.

\begin{figure}
\begin{center}
\includegraphics[scale=0.7]{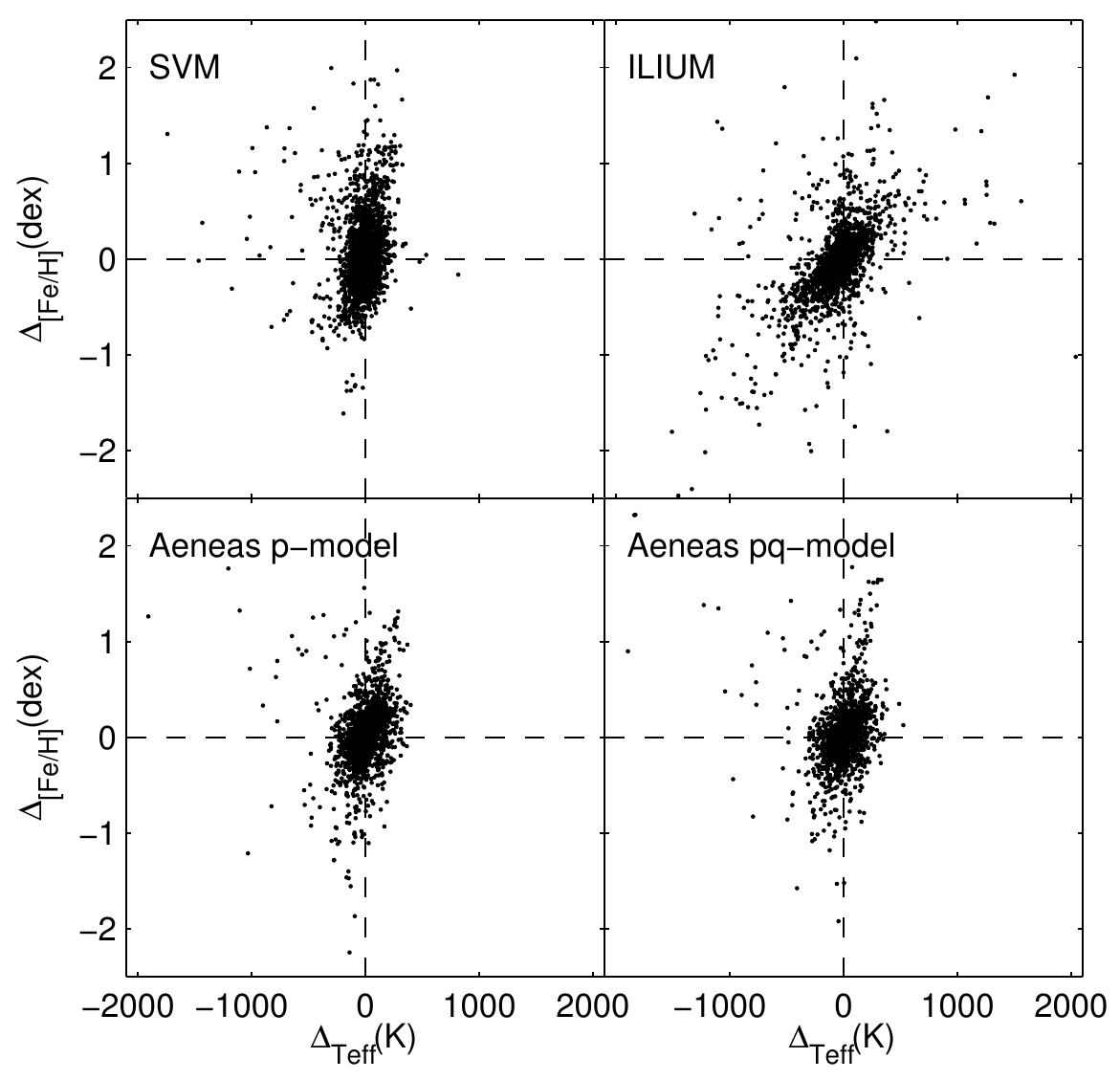}
\caption{The correlation between the $\teff$ and \feh\ residuals for stars with $G=15$\,mag for all four methods.}\label{fig:degfehteff}
\end{center}
\end{figure}

Figure~\ref{fig:degfehteff} plots the residuals in \feh\ against the residuals in $\teff$ for the four methods, and reveals a (weak) degeneracy between these parameters also. The reason is probably as follows: Either decreasing the metallicity of a star, or increasing its effective temperature, makes the spectrum bluer. To explain (to preserve) the observed colour of a star when we overestimate $\teff$, we must adopt a higher metallicity, i.e.\ we also overestimate \feh. Vice versa is also true. In other words, there is a positive $\teff$--\feh\ degeneracy. As with the $\teff$--$\A0$ degeneracy this is intrinsic to the data and not an artefact of the methods used. But unlike that degeneracy, the introduction of the HRD and parallax in pq-model does not help to significantly reduce the $\teff$--\feh\ degeneracy, as we can see if we compare the bottom two panels of Figure~\ref{fig:degfehteff}.

\subsection{$\feh$ estimation}\label{subsect:feh}

\begin{figure*}
\begin{center}
\includegraphics[width=15cm]{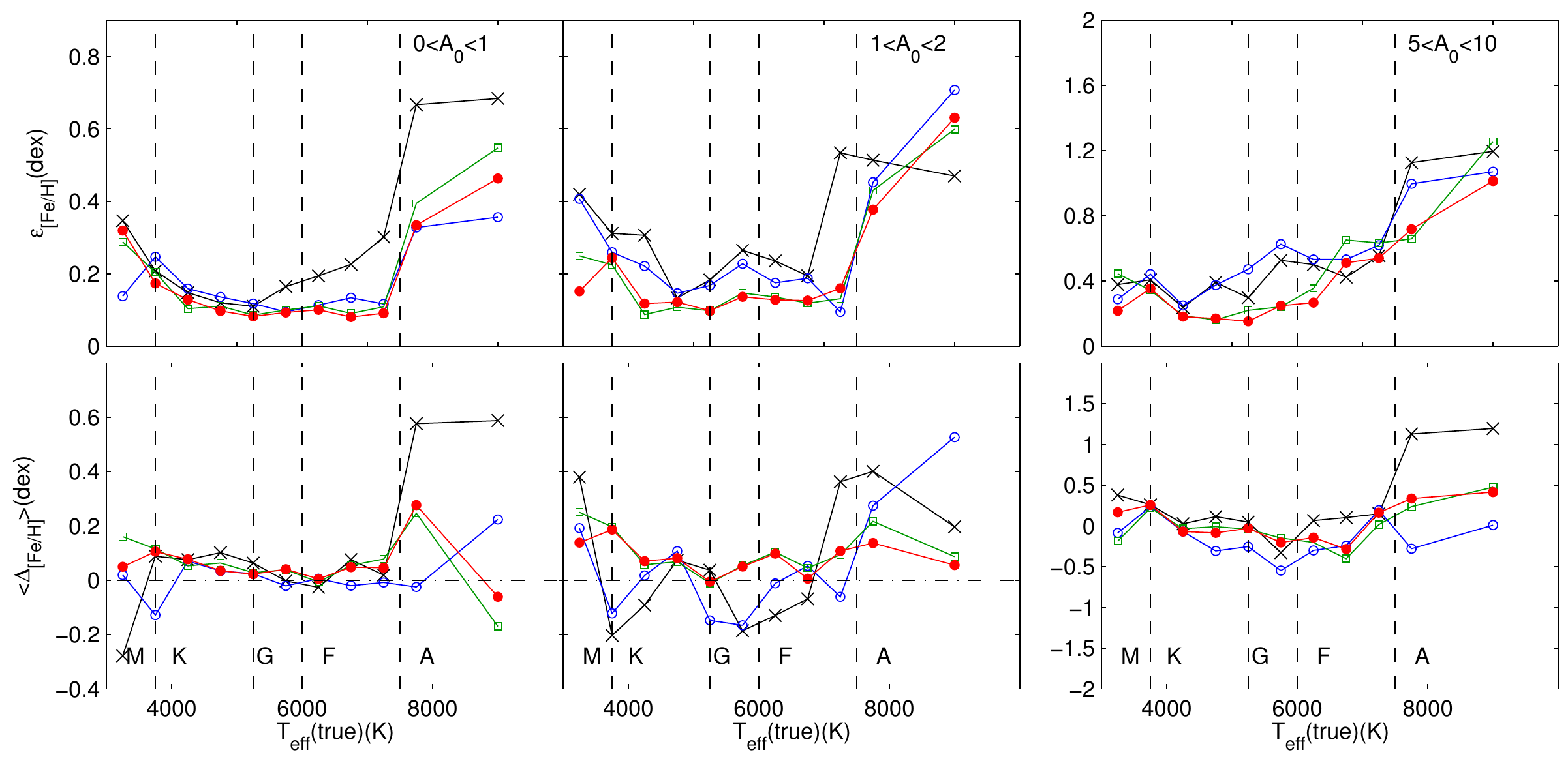}
\caption{The errors in estimated \feh\ as a function of true $\teff$ for stars with G=15\,mag. The top row shows the mean absolute residual, $\epsilon_{\rm \feh}$. The bottom row shows the mean residual, $\langle \Delta_{\rm \feh} \rangle$. The columns from left to right are for stars with $\A0<1$, $1<\A0<2$, and $5<\A0<10$, respectively. 
The colours/symbols indicate the method: SVM (black crosses), \ilium\ (blue open circles), Aeneas p-model (green rectangles), Aeneas pq-model (red closed circles).
The dashed vertical lines delineate the spectral types A, F, G, K, M. Note the change in scale for the highest extinction panels.}
\label{fig:feh}
\end{center}
\end{figure*}


The performance on metallicity (Figure~\ref{fig:feh}) is largely independent of $\teff$ for F,G,K stars, especially for Aeneas, but degrades for A and M stars. The lower accuracy for A stars is well established: it is due to the lack of metal-sensitive lines (visible at low resolution) forming in hotter photospheres. The (more modest) reduced accuracy for M stars is probably a consequence of the increased complexity of these spectra, in particular a \feh--$\logg$ and \feh--$\teff$ degeneracy. At the higher extinctions, performance degrades overall, as we would expect on account of the extra variance due to extinction, plus the $\teff$--$\A0$ degeneracy and the impact it has on metallicity (see previous section).

Overall the metallicity performance is good: 0.1\,dex for F,G,K stars for $\A0<2$\,mag with Aeneas.
Whereas \ilium\ was not very competitive in performance at $\teff$, here on metallicity it does well, at least at G=15. SVM shows the poorest performance overall, in particular for A,F,G stars at lower extinctions.

\subsection{$\logg$ estimation}\label{subsect:logg}

\begin{figure*}
\begin{center}
\includegraphics[width=15cm]{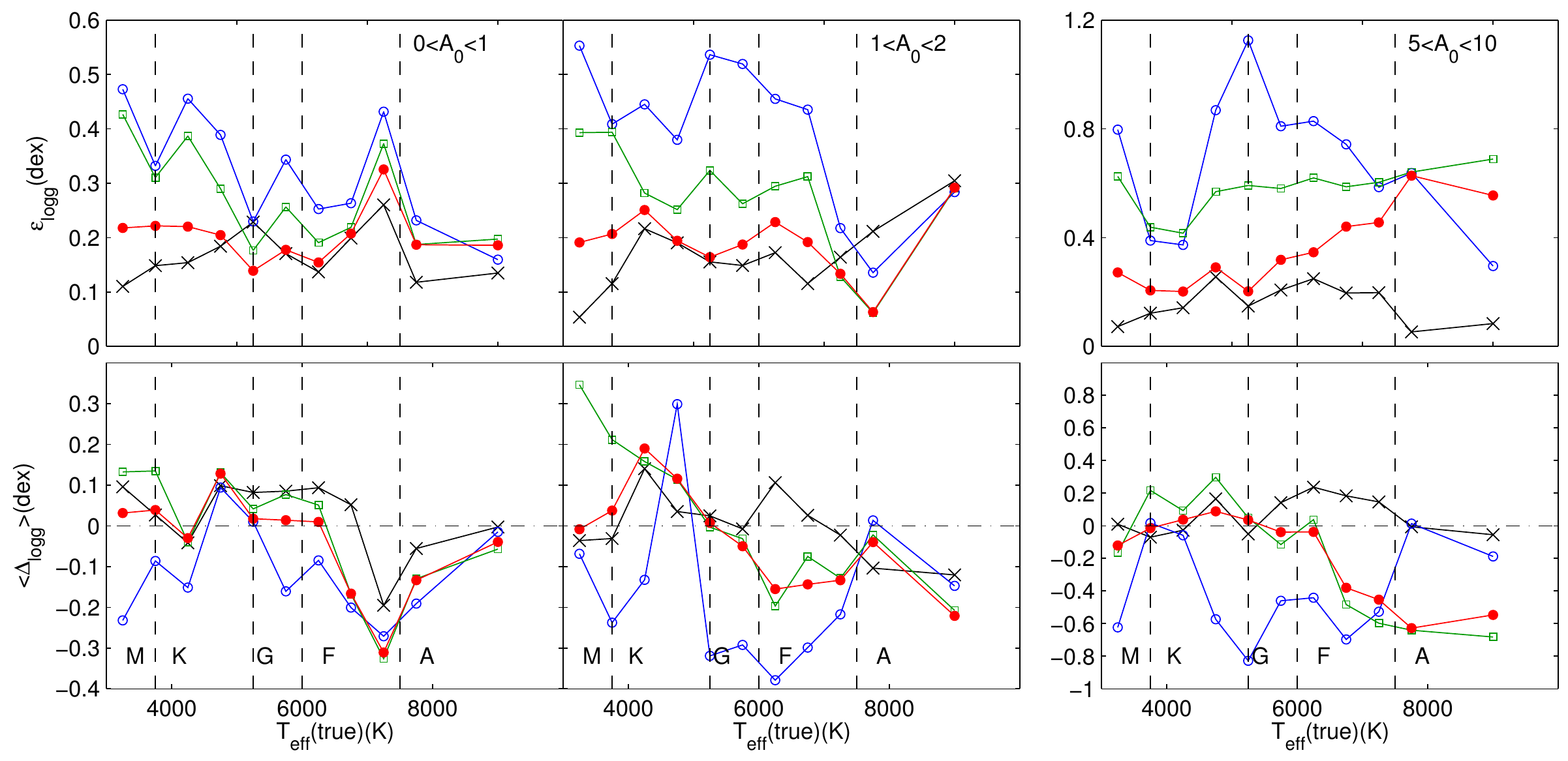}
\caption{The errors in estimated $\logg$ as a function of true $\teff$ for stars with G=15\,mag. The top row shows the mean absolute residual, $\epsilon_{\logg}$. The bottom row shows the mean residual, $\langle \Delta_{\logg} \rangle$. The columns from left to right are for stars with $\A0<1$, $1<\A0<2$, and $5<\A0<10$, respectively. 
The colours/symbols indicate the method: SVM (black crosses), \ilium\ (blue open circles), Aeneas p-model (green rectangles), Aeneas pq-model (red closed circles).
The dashed vertical lines delineate the spectral types A, F, G, K, M. Note the change in scale for the highest extinction panels.}
\label{fig:logg}
\end{center}
\end{figure*}

\begin{figure}
\begin{center}
\includegraphics[scale=0.7]{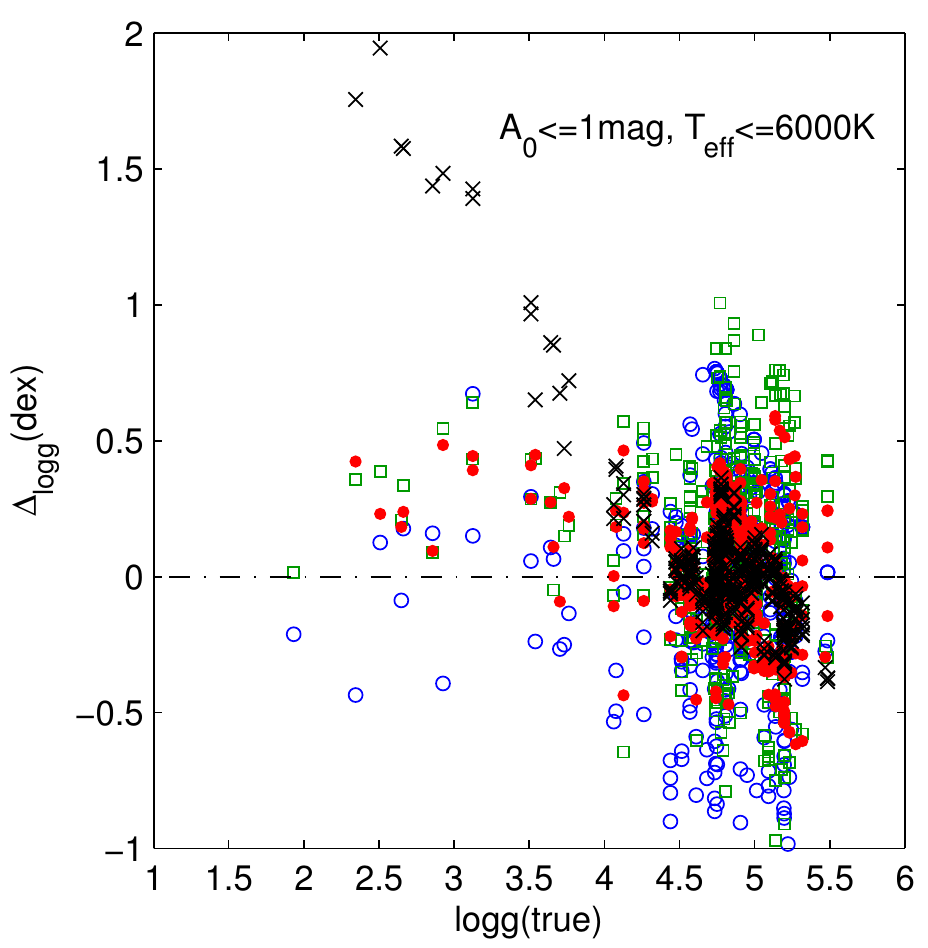}
\end{center}
\caption{The systematic bias of the estimated $\logg$ for the stars with $\A0<1$\,mag (G=15 test data set).
The colours/symbols indicate the method: SVM (black crosses), \ilium\ (blue open circles), Aeneas p-model (green rectangles), Aeneas pq-model (red closed circles).}\label{fig:loggbias}
\end{figure}

Surface gravity is generally the most difficult of the APs to estimate, as it is responsible for the least variance in the spectrum. Figure~\ref{fig:logg} shows the performance as a function of $\teff$.  Overall the accuracy is reasonable, around 0.1--0.3\,dex across all $\teff$ and $\A0$ for the overall best performing method, SVM.  For none of the methods is there a general dependence of the MAR on $\teff$, although there is degradation at the highest extinctions. The systematics can be considerable, and sometimes dominate the total errors, especially for \ilium.  Aeneas pq-model is better than p-model. This we would expect, because the combined use of the parallax, apparent magnitude and HRD helps to constrain $\MG$ which in turn helps to infer $\logg$. (More discussion of this is in section~\ref{subsect:ppqcompare}).

Figure~\ref{fig:loggbias} shows the correlation between $\Delta_{\logg}$ and the true $\logg$ for stars with $\A0<1$\,mag and $\teff<6000$\,K. Although SVM estimates $\logg$ accurately for dwarf stars, it significantly overestimates it for giant stars, much more so than the other methods. For a star with true $\logg\simeq2.5$\,dex, the systematic error is 1.5--2\,dex, which produces a dwarf surface gravity (4--4.5\,dex). Therefore, SVM is unreliable for cool giant stars. The good performance for SVM seen in  Figure~\ref{fig:logg} and Tables~\ref{tab:resultsg15} and~\ref{tab:resultsg19} is a result of the sample we are averaging over: there are far more dwarfs than giants in the test data set (see Figure~\ref{fig:apdist}), which may (partially) explain this poor performance on giants.
Aeneas also overestimates $\logg$ for giant stars, but by less than $0.5$\,dex, so a dwarf/giant separation to a few times the MAR is possible.  Although \ilium\ shows a larger dispersion in $\logg$ for the giants, it does not show any clear bias.

\section{Discussion}\label{sect:disc}

\subsection{AP accuracy vs.\ G magnitude}\label{subsect:APG}
\begin{figure*}
\begin{center}
\includegraphics[scale=0.7]{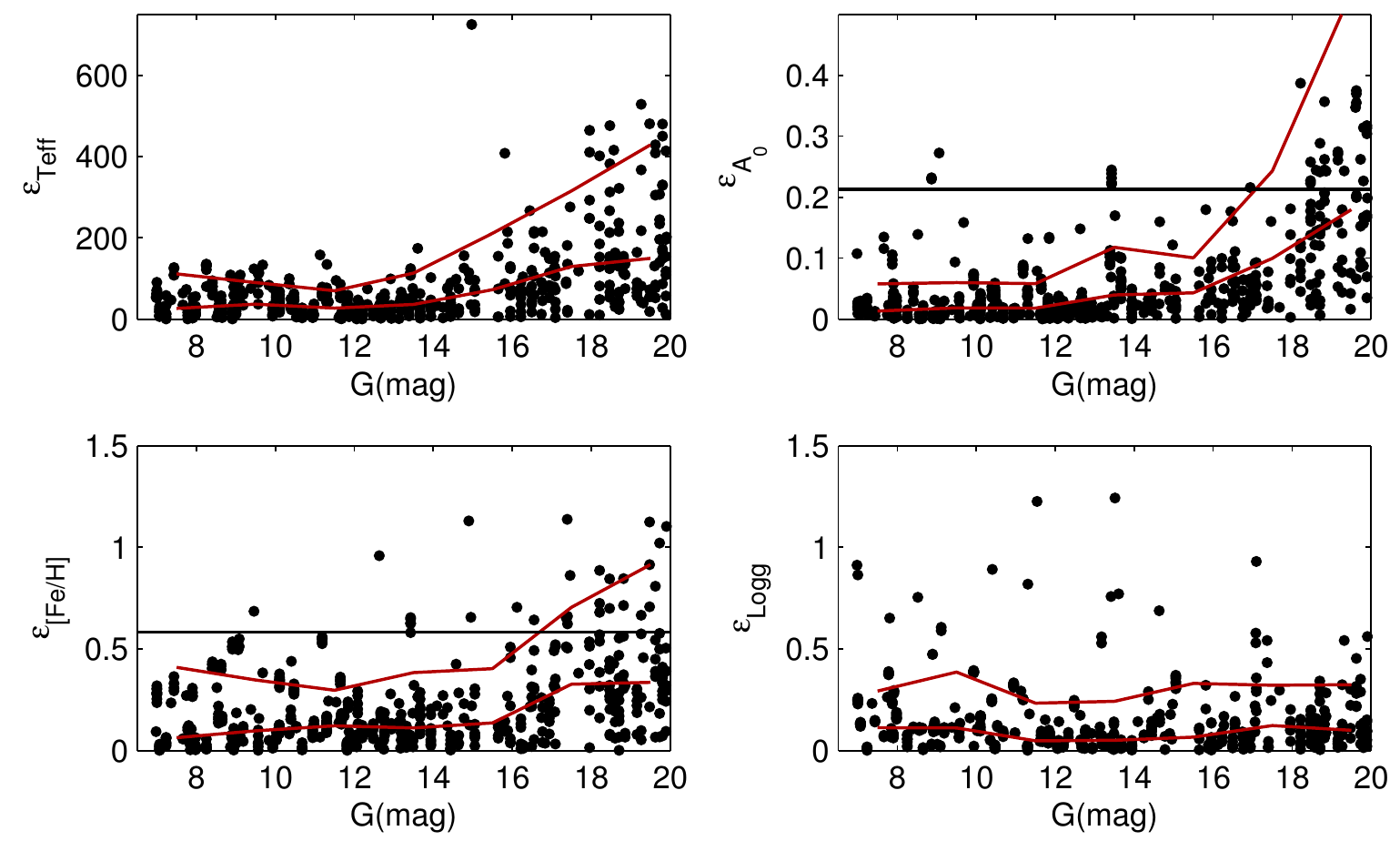} 
\caption{The mean absolute residual (MAR) of $\teff$, $\A0$, \feh, and $\logg$ for SVM as a function G magnitude. The black dots show $\epsilon_{\rm AP}$ for the individual stars with $\A0<1$\,mag, the red lines show the position of the 50\% and 90\% percentiles. The horizontal black lines show the MAR from doing \emph{prior-only} AP estimation 
based on the same training data set (namely, by assigning APs to the mean over the training data):
the two lines shown are $\A0=0.21$\,mag and \feh\,=\,0.58\,dex, while
the values for $\teff$ and $\logg$ are 2060\,K and 1.96\,dex respectively (off the scale of the plot).
These \emph{prior-only} values indicate how well we can do without using the data at all.}
\label{fig:performGsvm}
\end{center}
\end{figure*}

\begin{figure*}
\begin{center}
\includegraphics[scale=0.7]{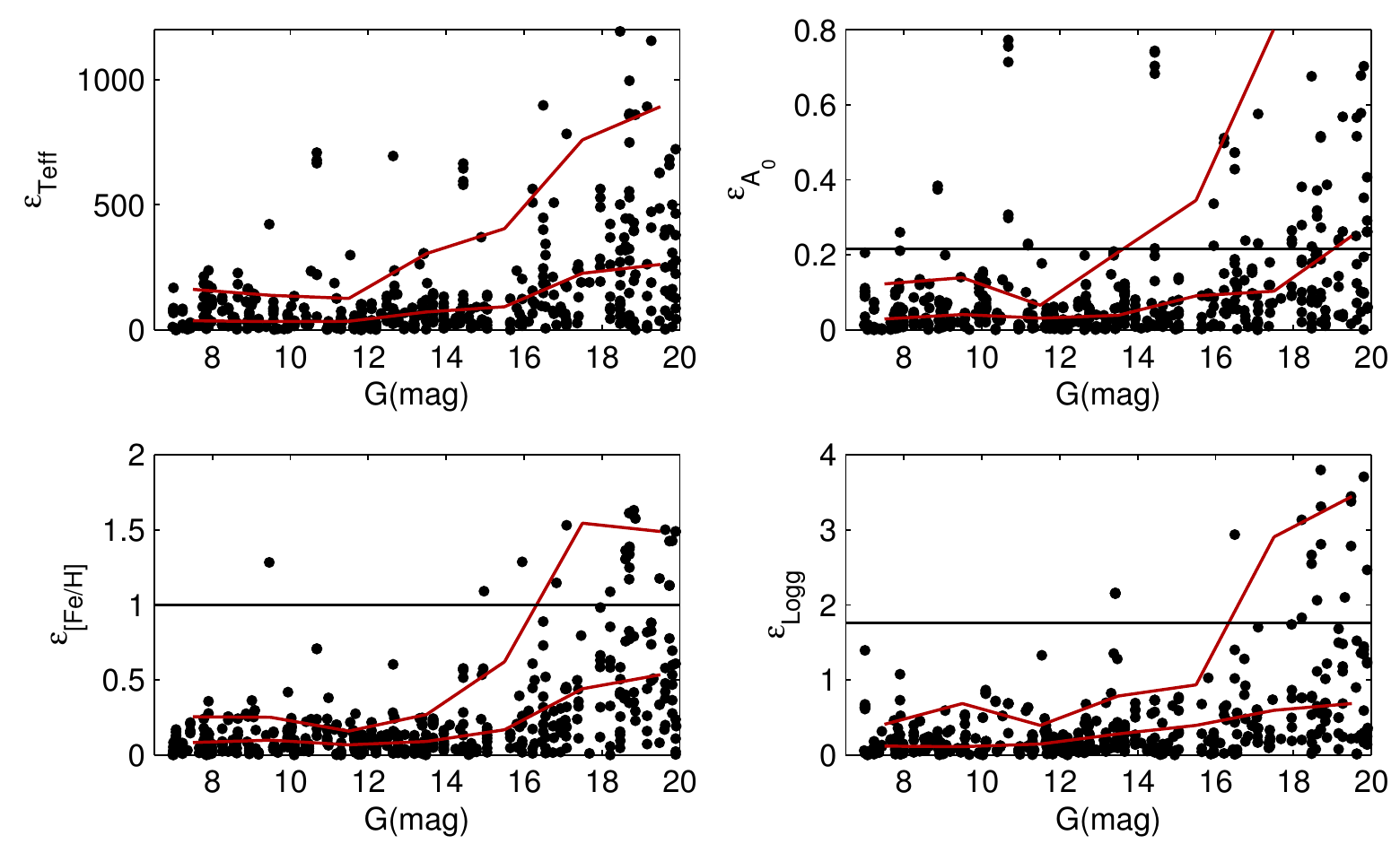} 
\caption{As Figure~\ref{fig:performGsvm}, but for \ilium.
The MAR for the \emph{prior-only} estimation are 
$\teff$\,=\,1860\,K, $\A0$\,=\,0.22\,mag, \feh\,=\,1.0\,dex, and $\logg$\,=\,1.76\,dex.}
\label{fig:performGilium}
\end{center}
\end{figure*}

\begin{figure*}
\begin{center}
\includegraphics[scale=0.7]{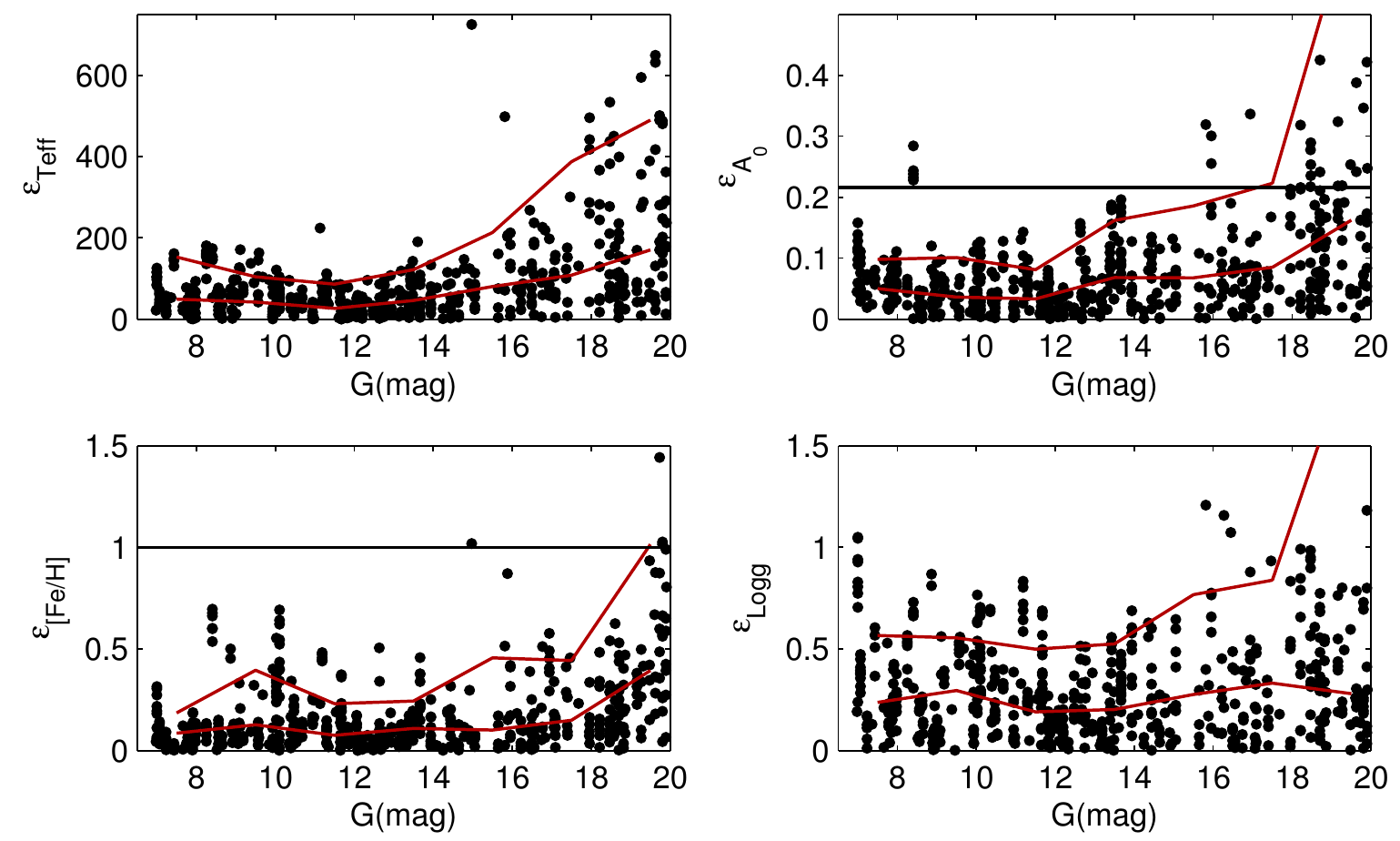} 
\caption{As Figure~\ref{fig:performGsvm}, but for Aeneas p-model.
The MAR for the \emph{prior-only} estimation are 
$\teff$\,=\,1860\,K, $\A0$\,=\,0.22\,mag, \feh\,=\,1.0\,dex, and $\logg$\,=\,1.76\,dex (the same as \ilium).}
\label{fig:performGpmodel}
\end{center}
\end{figure*}

\begin{figure*}
\begin{center}
\includegraphics[scale=0.7]{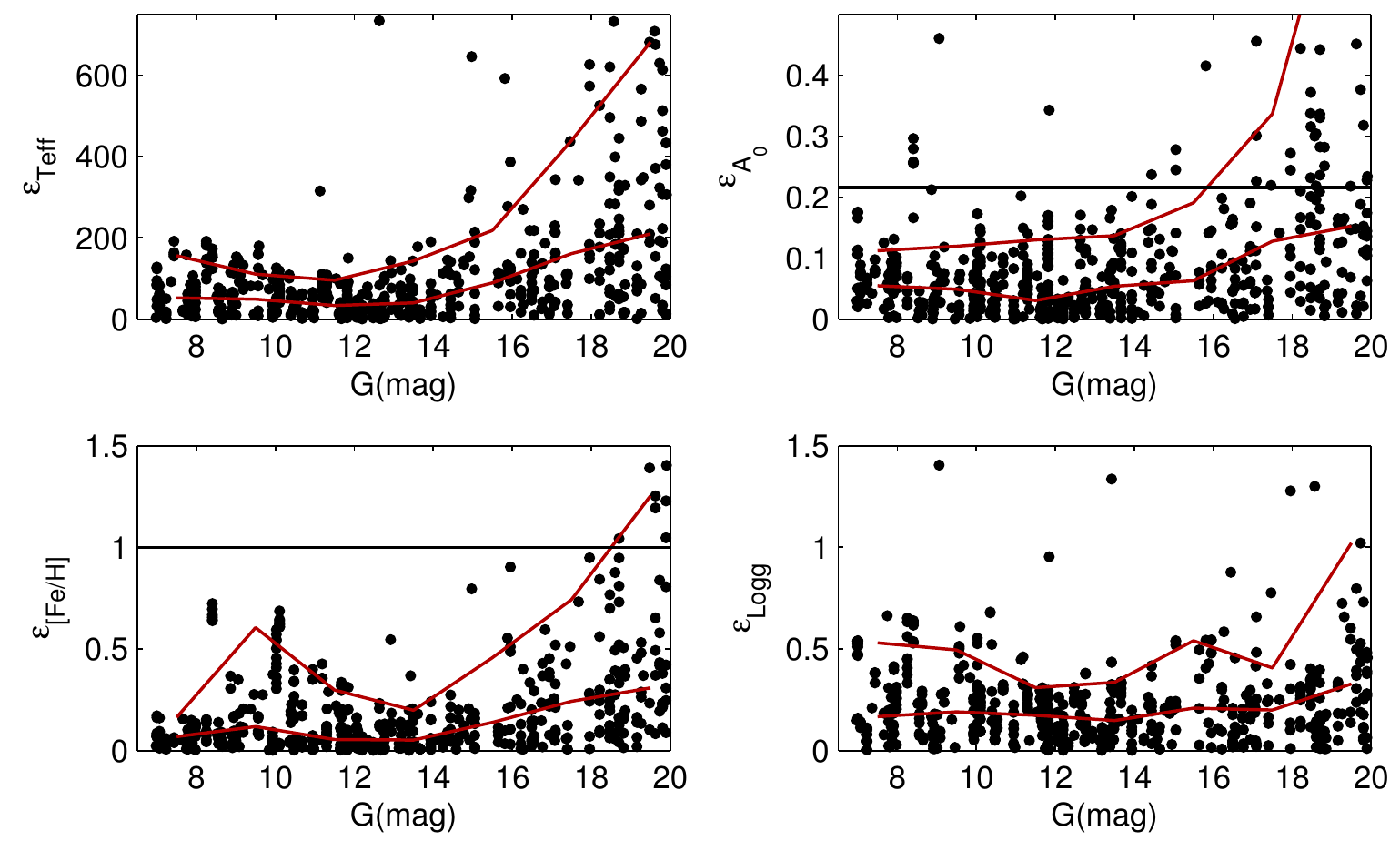} 
\caption{As Figure~\ref{fig:performGsvm}, but for Aeneas pq-model.
The MAR for the \emph{prior-only} estimation are 
$\teff$\,=\,1860\,K, $\A0$\,=\,0.22\,mag, \feh\,=\,1.0\,dex, and $\logg$\,=\,1.76\,dex (the same as \ilium).}
\label{fig:performGpqmodel}
\end{center}
\end{figure*}

We apply the methods to the mixed magnitude test set, 2000 stars with magnitudes ranging from G=6 to G=20. The MAR for each AP averaged over all APs for low extinction stars ($\A0<1$\,mag) is shown as a function of G for all four methods in Figures~\ref{fig:performGsvm} to~\ref{fig:performGpqmodel}. The performance starts to degrade between G=12 and G=16, depending on the method and AP. This is consistent with the magnitude range in which the BP/RP SNR begins to decrease due to the dominance of source photon noise over constant noise sources (background and detector).
It is instructive to compare these plots. For example, 
comparing Figures~\ref{fig:performGpmodel} and~\ref{fig:performGpqmodel}, we see how the introduction of the parallax/HRD into Aeneas improves the accuracy of $\logg$ across a range of magnitudes.

The performance here, as in the previous sections, of course depends on the distribution of APs in the data set.  Specifically, if the AP range of the test data set is constrained to some range (as it is both here and in reality), then just this (prior) knowledge represents an expected upper limit on the uncertainties in our AP estimates.  For example, most stars have \feh\ between -2.5 and +0.5\,dex, and if they were uniformly distributed over this range then the error from even a useless method (``guessing'') should not be more than 1.5\,dex on average.  We quantify the upper limit of a reasonable error as that calculated when we assign all stars an AP value equal to the mean of the training data set (mean-of-training-data model). We call this the \emph{prior-only} MAR. Strictly speaking it is only relevant when the training and test distributions are the same, but it gives us a useful baseline against which we can compare performance. These values are given in the figure captions of Figures~\ref{fig:performGsvm} to \ref{fig:performGpqmodel}, and plotted as black horizontal lines if they fit on the scale of the plot. Note that the value for $\A0$ is quite low because the sample is limited to stars with true $\A0<1$\,mag. For the full test data sets used before ($\A0$\,=\,0--10\,mag) it is 2.47\,mag for SVM and 3.26\,mag for \ilium\ and Aeneas.

We can also use this prior-only MAR to determine the G magnitude below which we can no longer usefully estimate the APs from the data. With SVM we can obtain useful estimates for $\teff$ and $\logg$ right down to G=20 (Figure~\ref{fig:performGsvm}). For $\A0$ and \feh\ this is not the case, with 10--50\% of stars with $G>17$ having higher errors from SVM than from the mean-of-training-data model. These statements only apply to low extinction stars, which explains the apparently bright magnitude limit for $\A0$. (If we extend to higher extinctions, then the prior-only MAR is larger, so the magnitude limit for useful estimates is fainter.) A similar approach can be taken for the other algorithms (which use a different training data set so have different prior-only MARs).  \ilium\ is again ``useful'' for all $\teff$, but for G$>$19 more than half of the stars have residual larger than the prior-only MAR.  Aeneas is useful for all APs over the whole range, although at the Gaia magnitude limit not much less than half of stars have residuals larger than the prior-only MAR. (Recall, however, that by taking the {\em mean} of the posterior PDF in Aeneas, we may be limiting the accuracy for low extinction stars).

\subsection{Aeneas p-model vs.\ pq-model}\label{subsect:ppqcompare}
\begin{figure}
\begin{center}
\includegraphics[scale=0.6]{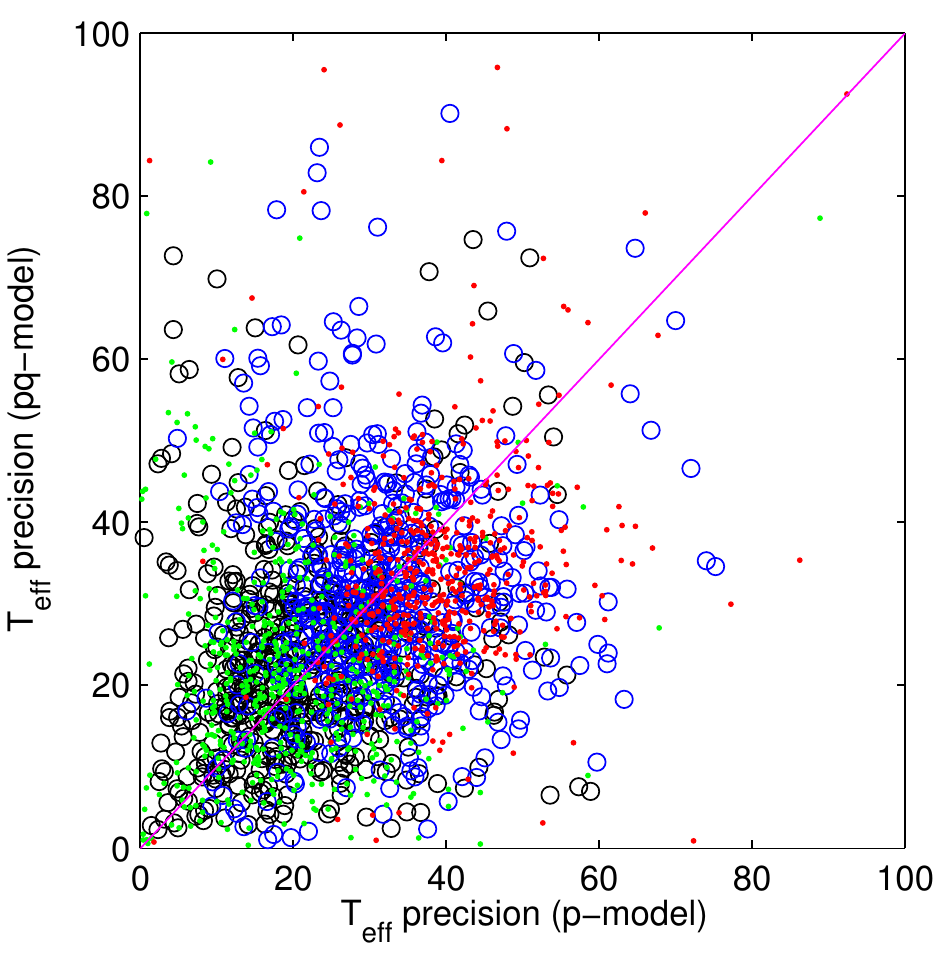}
\includegraphics[scale=0.6]{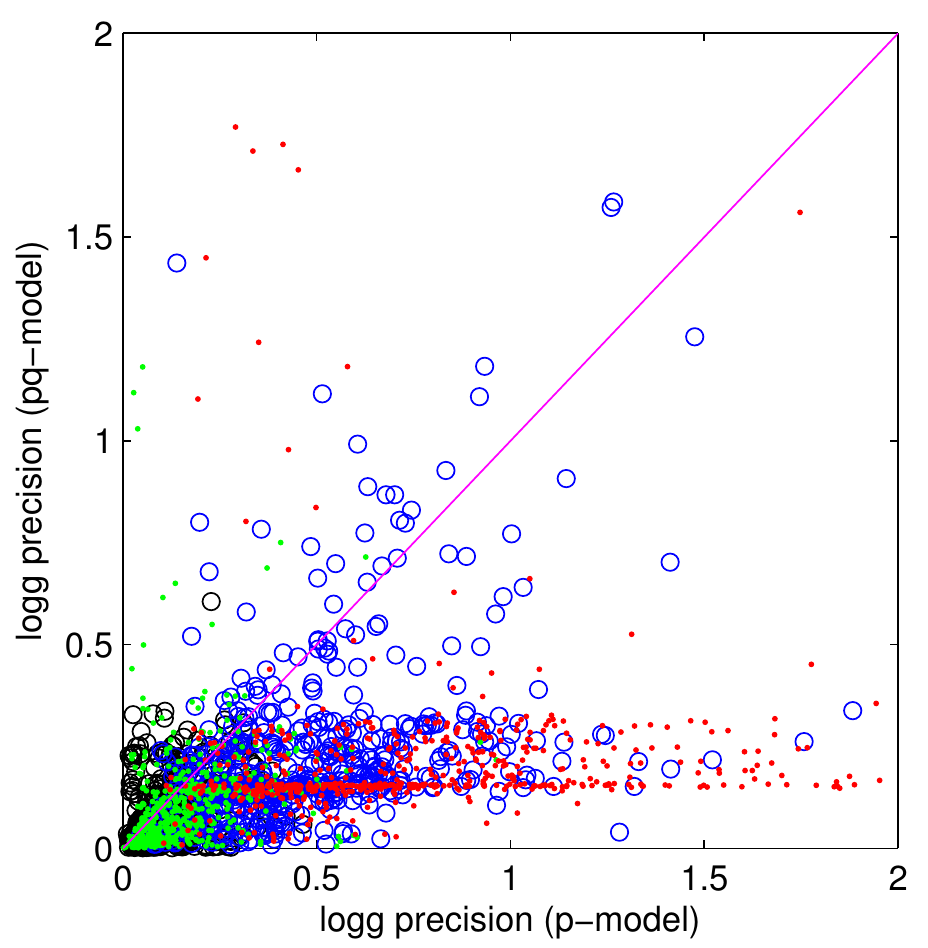}
\end{center}
\caption{The precision in $\teff$ (top panel) and $\logg$ (bottom panel) from Aeneas p-model vs.\ pq-model. 
The precision is defined as half the interval which encloses 68\% of the posterior probability (equal to 1$\sigma$ if the distributions were Gaussian) for the projected one-dimensional posterior PDFs from Aeneas.
Stars with $\A0<1$\,mag are plotted with black circles (for G=15) and blue circles (G=19).
Stars with $\A0>5$\,mag are plotted with green dots (for G=15) and red dots (G=19).}
\label{fig:loggerraeneas}
\end{figure}

\begin{figure*}
\begin{center}
\includegraphics[scale=0.55]{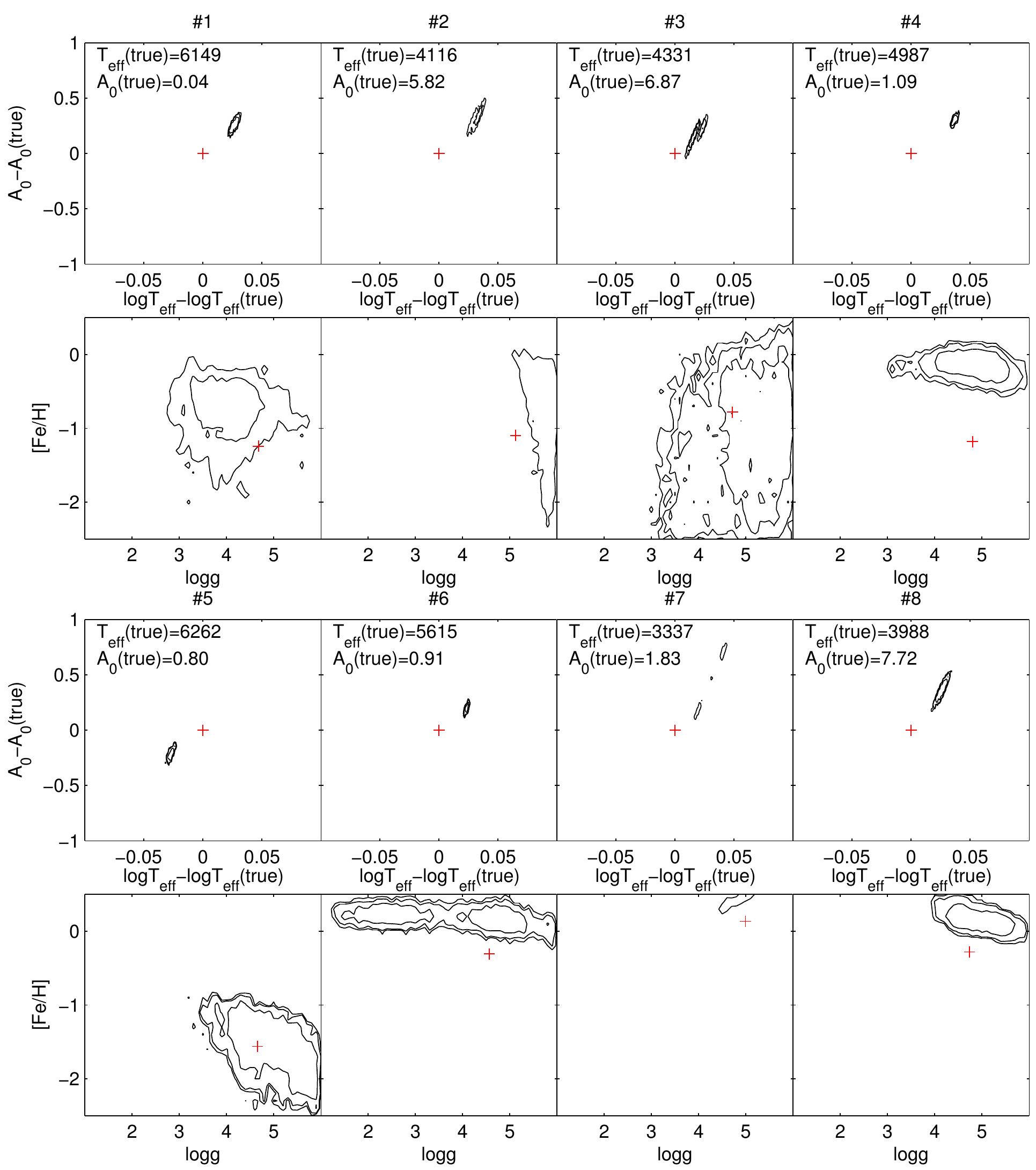}
\caption{4D Posterior PDF for Aeneas p-model plotted as two 2D projections for each of a sample of eight stars at G=19. The contours enclose 95\%, 90\% and 68\% of the probability. The red cross shows the true APs. The $\teff$--$\A0$ plots are centred on the true APs, and the axes are the AP values relative to these.}\label{fig:postp-p}
\end{center}
\end{figure*}

\begin{figure*}
\begin{center}
\includegraphics[scale=0.55]{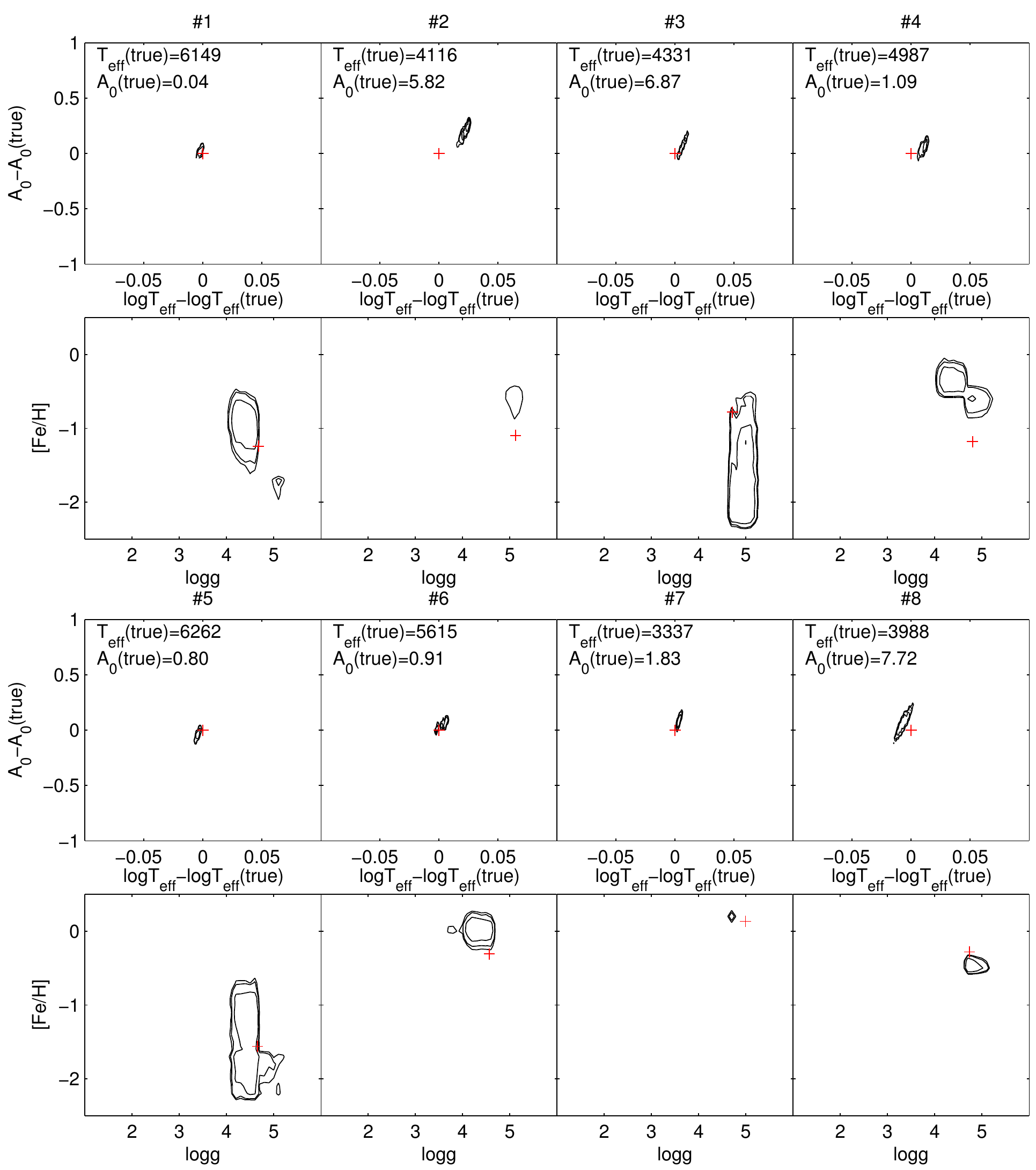}
\caption{As Figure~\ref{fig:postp-p}, but now for pq-model.}\label{fig:postp-pq}
\end{center}
\end{figure*}

It is of interest to compare the results from p-model and pq-model and to identify the role of the parallax and the HRD prior in AP estimation.  

At brighter magnitudes (e.g.\ G=15), the pq-model does not perform significantly better than the p-model on $\teff$, $\A0$, and \feh.
 However, the pq-model does produce considerably more accurate $\logg$ estimates, as can be seen in Figures~\ref{fig:histres} and~\ref{fig:logg}.  The latter shows noticeable improvement at all extinction ranges, in particular for cooler stars.
(At G=19 the performance of pq-model relative to p-model on $\logg$ is ever better; see Table~\ref{tab:resultsg19}.)

We also find that the posterior PDF from the pq-model is generally narrower than that from the p-model.  This width is important, because it is a measure of the precision of (confidence in) our AP estimates. (Note the distinction between ``precision'' and ``accuracy'', the latter referring to the size of the residuals.)  This can be seen in the top panel of Figure~\ref{fig:loggerraeneas}, which plots the projected width of the posterior PDF from the pq-model against that from the p-model, for individual stars.  Width is defined here as half the interval which encloses 68\% of the probability (equal to 1$\sigma$ if the distributions were Gaussian).  The precision of pq-model estimates are better for faint stars (red points) in particular.  The lower panel of this plot shows a more pronounced difference in the precisions of pq-model and p-model for $\logg$, at both G=15 (green points) and G=19 (red points), and also for low extinction stars (blue circles), than for $\teff$. 
In other words, the information provided by the parallax in conjunction with the HRD (which leads to information on 
$\MG$) {\em given} that we have the BP/RP spectrum, has more impact on improving the $\logg$ precision (and accuracy, as we have seen) than the $\teff$ precision and accuracy. (We may not say -- as often is said -- that the parallax simply ``tells us'' $\logg$.)

Figures~\ref{fig:postp-p} and~\ref{fig:postp-pq} show a few examples of the posterior PDF for eight stars at G=19\,mag (each 4D PDF is projected into two 2D plots).  For the weak parameters ($\logg$ and $\feh$), the PDF shrinks considerably when we introduce the parallax/HRD. Overall, we find that the parallax/HRD helps to reduce the uncertainty of the estimated APs, in particular for faint stars, and especially for the weak APs. As we have seen in earlier sections, in some parts of the AP space it also improves the accuracy of the AP estimations (i.e.\ produces a lower MAR). We note, however, that in particular for the strong APs, Aeneas tends to underestimate the uncertainties (PDF width too small). This is being investigated.

In principle pq-model should never be worse than p-model. This is because as long as the HRD prior used in the pq-model is consistent with the test data, then in the limit of a very poor parallax the HRD should simply add no information. We see from Tables~\ref{tab:resultsg15} and~\ref{tab:resultsg19} that this is true at G=15, but not at G=19. That is because at fainter magnitudes we have much larger fractional parallax errors -- both the parallax error is larger and the stars are more distant -- and so the parameter $q$ is far noisier. This affects intrinsically bright stars -- upper main sequence stars and giant stars -- in particular. Although in principle the error model for $q$ should accommodate this, we expect that the Gaussian approximation we use will become increasingly poor for larger fractional parallax errors. Furthermore, as discussed in section~\ref{sect:grids} we do have inconsistencies between the HRD and the test data set, and these will have a more significant negative impact on poorer data (as we then rely more on the prior). 
It will be necessary to improve this before the full benefit of the pq-model at faint magnitudes can be realized.

The differences between p-model and pq-model were also explored in CBJ11, where Aeneas was applied to Hipparcos/2MASS data.  Section 3.7 of that paper also gives some intuitive insight into how the model uses the information in the HRD and parallax.

\subsection{Comparison with SSPP}

\begin{table} 
\caption{Comparison of the AP estimation accuracy on Gaia BP/RP spectra from this paper with the AP estimation accuracy on SDSS/SEGUE spectra from the SSPP. All values given are the RMS of the residuals. Note that the AP and SNR distributions in the samples are very different, so the comparison should not be over-interpreted.}
\label{tab:snrG}
\begin{center}
\begin{tabular}{lccccc}
\hline
AP & SVM & \ilium\  & Aeneas & Aeneas & SSPP\\
 & & & p-model & pq-model \\
\hline
$\teff$\,/\,K        & 98 & 198 & 110 & 119 & 57\\
\feh\,/\,dex      & 0.25 & 0.27 & 0.19 & 0.21 & 0.10\\
$\logg$\,/\,dex & 0.26 & 0.60 & 0.37 & 0.28 & 0.17\\
\hline\end{tabular}
\end{center}
\end{table}%
It is informative to compare the overall AP estimation accuracy on the (simulated) BP/RP spectra with what has been achieved on real spectra from spectroscopic surveys. (Those surveys of course also rely ultimately on synthetic model spectra in order to infer physical parameters.)  \citet{lee08} apply a number of methods to estimate APs from SDSS/SEGUE spectra (in the processing system called the SEGUE Stellar Parameter Pipeline, SSPP).  The variation of SNR with wavelength for the SDSS spectra is quite different from that in the BP/RP spectra -- not least because of the highly nonlinear dispersion of the latter -- making a comparison difficult.  But for a rough idea, we compare the performance of stars from SSPP with SNR$>$20 per pixel, which have SDSS g-band magnitudes between 17.5 and 19.5, with our results for G\,$>$\,18.5. (The SNR of the BP/RP spectra at G=19 is shown in Figure~\ref{fig:snr_G19}).  As most SSPP stars are expected to have low extinction (SSPP does not estimate $\A0$), we use our results for $\A0<1$. Table~\ref{tab:snrG} compares the numbers.
Given the much higher resolution of the SEGUE spectra (by a factor of about 100), it is not surprising that SSPP gives better accuracy in all cases, although not by a large margin. Recall also that our methods have been trained to cover a very wide range  of interstellar extinction (unlike the SSPP methods), yet the comparison here is for a sample dominated by low extinction stars.

\subsection{Science cases}

So far we have analysed the AP accuracy by averaging over a broad range of APs, even when analysing trends with $\teff$ or G and/or restricting the analysis to narrower extinction ranges. Yet certain categories of stars -- ranges of APs -- are of particular interest for addressing the Gaia science cases on Galactic structure and stelar populations.  In practice we must of course identify these stars using their estimated, rather than true, APs.  We analyse here three particular science cases, each corresponding to a particular type of star, selecting them based on their estimated APs .
We use the mixed magnitude test data set (G=6--20).
As the selected samples turned out to be rather small, we created a new test set with 20\,000 rather than 2000 stars, selected in the same way (same AP distribution), in order to increase the statistical significance of the results.

The first science case concerns F/G dwarfs, samples of which are used to study the kinematics of the Galactic disk(s) and halo. It is important to be able to select these with high completeness and low contamination, although in general increasing one metric decreases the other (see \citealt{cbj08} for a related study in Gaia on quasar/galaxy/star sample separation). Using the mixed magnitude test set with true $\A0<1$\,mag, we select F/G dwarfs according to the estimated effective temperature and surface gravity: we define them as objects with $5250<\teff<7500$ and $\logg>4.5$. Given that there are 1629 true objects falling into these ranges, we can calculate the sample completeness (fraction of true positives) and sample contamination (fraction of false positives) for each AP estimation method. The results are in Table~\ref{tab:fgdwarfs}. SVM achieves the highest completeness, but also the highest contamination. \ilium\ achieves the lowest of both, and the two Aeneas models achieve a (arguably preferable) compromise between completeness and contamination. The MARs from Aeneas pq-model of $\teff$, $\A0$, $\feh$ and $\logg$ for the stars in the sample which are truly F/G stars are 108\,K, 0.08\,mag, 0.15\,dex, and 0.18\,dex, respectively. This is good accuracy in \feh, for example, for doing chemodynamical studies with what will be an enormous sample. Note that if we also take into account the AP estimation precision (provided naturally by \ilium\ and Aeneas), and/or take a narrower AP range to define F/G dwarfs, then we could decrease the sample contamination at the cost of a lower sample completeness.
\begin{table}
\caption{The performance of the selection of F/G dwarfs (mixed magnitude test data set for $\A0<1$\,mag) by the four methods. There are 1629 true F/G dwarfs.}\label{tab:fgdwarfs}
\begin{tabular}{ccccc}
\hline
&  SVM & \ilium\ & Aeneas & Aeneas \\ 
& & & p-model & pq-model \\
\hline
Selected stars &    1886	& 1084  & 1462	& 1493 \\
completeness &	  93\%	 & 62\%	& 82\%	& 82\% \\
contamination &  20\%	 & 7\%	& 9\%     & 10\% \\
\hline
\end{tabular}
\end{table}

The second science case is K giants. Being bright, they can be selected out to tens of kpc from the Sun and so used to study the halo. However, it is hard to select K giants from multi-band photometry. 
We define K giants as stars with $\teff<5250$ and $\logg<4$. Limiting ourselves to stars with $\A0<1$\,mag (we want to study the halo), we find just 99 true objects (a consequence of having used the HRD to build the test data sets). Proceeding as before, we find the sample statistics reported in Table~\ref{tab:kgaints}. SVM finds no stars at all. Its poor performance on cool giants was already demonstrated in section~\ref{subsect:logg}.
Conversely \ilium\ finds 80\% of them, but at the cost of
a sample in which three quarters are not K giants. 
Aeneas pq-model arguably does the best over all, finding a sample of 80, of which 74 are K giants and 6 are not.
The MARs of $\teff$, $\A0$, $\feh$ and $\logg$ for these 74 from Aeneas pq-model are 111\,K, 0.15\,mag, 0.16\,dex, and 0.50\,dex, respectively.
\begin{table}
\caption{The performance of the selection of K giants (mixed magnitude test data set for $\A0<1$\,mag) by the four methods. There are 99 true K giants.}\label{tab:kgaints}
\begin{tabular}{ccccc}
\hline
&  SVM & \ilium\ & Aeneas & Aeneas \\
& & & p-model & pq-model \\
\hline
Selected stars & 0	& 313	& 134	& 80\\
completeness &0\%	& 80\%	& 90\%	& 75\%\\ 
contamination &0\% & 75\%	& 34\%	& 8\%\\
\hline
\end{tabular}
\end{table}

The final science case concerns selecting A stars, a useful tracer of the Galactic disk (amongst other things) where they can be used to construct a 3D model of the interstellar extinction.
A stars are defined here as any star with $\teff>7500$\,K, regardless of the value of the other three APs. The completeness and the contamination of the samples selected by the four models are listed in Table~\ref{tab:astars}, which separates the statistics into $\A0<1$, $3<\A0<5$ and $5<\A0<10$ (selected on the true extinction). For $\A0<1$\,mag
we can build a sample of A stars which is 95\% complete with 5\% contamination, using any of the methods. At higher extinctions completeness drops, but perhaps surprisingly, at the highest extinctions a low contamination (8\%) is possible with SVM or Aeneas.

\begin{table}
\caption{The performance of the selection of A stars (mixed magnitude test data set for $\A0<1$\,mag) by the four methods. $N$ gives the number of true A stars in each extinction range.}\label{tab:astars}
\begin{tabular}{ccccc}
\hline
&  SVM & \ilium\ & Aeneas & Aeneas \\
& & & p-model & pq-model \\
\hline
\multicolumn{5}{c}{$\A0<1$, N=264}\\\hline
Selected stars&	264	&261   &	264	&266\\
completeness&	96\%& 94\% & 95\%      &95\%\\
contamination &        4\%& 5\%   & 5\%     &6\%\\
\hline
\multicolumn{5}{c}{$3<\A0<5$, N=137}\\\hline
Selected stars &       157	& 159 &	135 &	132\\
completeness&	84\%& 78\%&	81\% &80\%\\
contamination&	27\%&33\%& 18\% &17\%\\
\hline
\multicolumn{5}{c}{$5<\A0<10$, N=196}\\\hline
Selected stars &	130	&259	&150	&155\\
completeness&	61\%&77\%      &71\%   & 69\%\\
contamination&	8\%	&42\%	&7\%	&12\%\\
\hline
\end{tabular}
\end{table}

In summary, for the F/G dwarf and A star science cases, all four methods give reasonable performance in terms of target selection, although \ilium\ is slightly inferior. In the K giant science case, Aeneas gives good completeness but still a rather high contamination. SVM is no use here (nothing found), because it has problems estimating $\logg$ for giants. This may be related to the extreme paucity of giants in the SVM training data set, such that SVM hardly learns to recognize them.

\section{Conclusions}\label{sect:summ}

We have used three algorithms, SVM, \ilium, and Aeneas (p-model and pq-model) on simulated Gaia spectrophotometry (BP/RP) to investigate how accurately we will be able to estimate stellar astrophysical parameters on the real Gaia data. These algorithms make up GSP-Phot, that part of the DPAC processing system which will estimate APs for all $10^9$ stars which Gaia will observe.

SVM achieves good overall performance for both the ``strong'' APs ($\teff$ and $\A0$) and the weak APs (\feh\ and $\logg$). 
A closer investigation in section~\ref{subsect:logg} showed that while SVM achieves accurate results for $\logg$ on dwarfs, it significantly overestimates $\logg$ for giants, making it unsuitable for selecting samples of giant stars. This may be an issue with the AP distribution in the training data set.  We also found that when the extinction is high ($\A0>5$\,mag), the accuracy of SVM on $\teff$ and $\A0$ degrades due to the $\teff$--$\A0$ degeneracy.
 
\ilium\ works well for the brighter stars (high SNR), but the performance degrades significantly for fainter stars, something which was also found in CBJ10. Despite this, \ilium\ frequently provides the least biased estimation among the four methods (e.g.\ Fig.~\ref{fig:loggbias}).  \ilium\ (and Aeneas) has the advantage over SVM that its performance is less sensitive to the distribution of the APs in the training grid: SVM suffers when this distribution is significantly different from the true (and unknown) distribution in the target data set. Experience has also shown that SVM performs better when trained on the random grid than on the regular grid. Yet because the random grid is produced by interpolating a regular grid, SVM involves two (noisy) interpolations (the second being the training of SVM). \ilium\ was motivated, in part, to avoid this, as the forward model can be fitted using a regular grid with little dependence on the exact AP distribution.

Although in Tables~\ref{tab:resultsg15} and~\ref{tab:resultsg19} Aeneas does not show smaller MARs than SVM, it
does show better performance in a number of other important parts of the AP space, both in terms of smaller MAR (higher accuracy) and smaller MR (less bias). 
(One should always remember that the performance summary obtained from averaging over a wide AP range depends sensitively on the AP distribution in the test set.)
Because it infers the posterior PDF over APs, Aeneas also has the advantage of naturally reporting AP uncertainties and of characterizing degeneracies or multimodality. In many respects Aeneas is the preferred algorithm, whether we use just the spectrum (p-model) or also include the HRD and parallax (pq-model). We saw that introducing the HRD and parallax considerably improved the accuracy of estimation of $\logg$ for both dwarfs and giants. However, there remains an issue with using pq-model at faint magnitudes 
(related in part to the inconsistencies discussed in section~\ref{sect:grids})
which needs to be resolved.
Further improvements to Aeneas are also possible, which together should result in the use of HRD and parallax improving the AP accuracy further.
Another issue with Aeneas is that it is considerably slower than \ilium\ or, in particular, SVM.
Work continues to accelerate it, and it may well be feasible to apply it to all $10^9$ stars which Gaia observes.

The current plan is for Gaia to accumulate data for five years after its launch in late 2013. Allowing time for post-mission processing, a final release of the full catalogue is expected in around 2021. There will, however, be earlier partial data releases, which will include both the BP/RP spectra and the corresponding APs from GSP-Phot. Details will be available on the Gaia website in due course.

This article is not the final word on stellar AP estimation with Gaia. As mentioned in the introduction, there are several other algorithms in the Gaia data processing system which deal with specific types of stars (e.g.\ hot stars, emission line stars) and with the higher resolution RVS spectrum, although only GSP-Phot provides APs for all stars. But even within the context of the suite of algorithms presented here for ``normal'' stars, there is room for improvement.  The obvious next step is to improve the consistency of the spectral grids with the HRD for use in Aeneas pq-model. Improvements could also be made in the random grids, particularly in the AP distribution, which needs to be widened in $\teff$. (GSP-Phot has been used on hotter stars, but there are issues with making the different spectral libraries consistent across the wider $\teff$ range, so those results have been omitted for now.) Here we only reported on estimates for the four main APs, but preliminary work has shown that we may also be able to estimate the parameter in the extinction law, $\R0$. Extension to an alpha element abundance parameter, [$\alpha$/Fe], may be difficult.  We also want to investigate combining the algorithms.  For instance, we could first estimate the strong APs with SVM, fix those, and then use \ilium\ to estimate the weak APs only. No doubt further improvements and modifications will become apparent when we finally receive real Gaia data.

\section*{Acknowledgments}

We thank our Gaia colleagues at MPIA -- Rene Andrae, Kester Smith, and Vivi Tsalmantza -- for thought-provoking discussions, and Kester Smith for assistance in assembling the data sets.  Ron Drimmel emphasized the convenience of using the extinction parameter, $\A0$, as an AP in place of the extinction in a band, $\AG$, something adopted here. This work was supported by a grant from the German space agency, DLR, as well as by
ASI contract I/058/10/0 and the MICINN (Spanish Ministry of Science and Innovation) - FEDER through grant AYA2009-14648-C02-01 and CONSOLIDER CSD2007-00050.  The GOG simulations were run on the supercomputer {\em MareNostrum} at the Barcelona Supercomputing Center -- Centro Nacional de Supercomputaci\'on.

\appendix
\section{On-line tables of the true and estimated APs of the test data}

\begin{table}
\begin{center}
\caption{The results of the application of the four methods to the 2000 stars in the $G=15$ data set defined in section~\ref{sect:grids}. The columns are defined in Table~\ref{tab:datacols}. The table is too wide to include here so is available in the online version.}
\label{tab:datag15}
\end{center}
\end{table}

\begin{table}
\begin{center}
\caption{The results of the application of the four methods to the 2000 stars in the $G=19$ data set defined in section~\ref{sect:grids}. The columns are defined in Table~\ref{tab:datacols}. The table is too wide to include here so is available in the online version.}
\label{tab:datag19}
\end{center}
\end{table}

\begin{table}
\begin{center}
\caption{The results of the application of the four methods to the 2000 stars in the mixed magnitude (G=6--20) data set defined in section~\ref{sect:grids}. The columns are defined in Table~\ref{tab:datacols}. The table is too wide to include here so is available in the online version.}
\label{tab:datagmixed}
\end{center}
\end{table}

\begin{table}
\begin{center}
\caption{The definition of the columns in the three online data
tables. Table~\ref{tab:notation} defines the notation and units, the
exception being that here the parallax is in mas and not arcseconds. 
$\sigma()$ denotes the approximate $1\sigma$ precision in the AP
estimates. All quantities labelled ``true'' are noise-free; they have
not been derived by the AP estimation methods.}
\label{tab:datacols}
\begin{tabular}{ll}
\hline
     1	 & ID \\
     2	 & G (true) \\
     3	 & $\teff$ (true) \\
     4	 & $\A0$ (true) \\
     5	 & $\logg$ (true) \\
     6	 & \feh\ (true) \\
     7	 & $\MG$ (true) \\
     8	 & $\AG$ (true) \\
     9	 & $\parallax$ / mas (true) \\
    10         & $\sigma(\parallax)$ / mas (true) \\
    11	 & $\teff$ (SVM) \\
    12	 & $\sigma(\teff)$ (SVM) \\
    13	 & $\A0$ (SVM) \\
    14	 & $\sigma(\A0)$ (SVM) \\
    15	 & $\logg$ (SVM) \\
    16	 & $\sigma(\logg)$ (SVM) \\
    17	 & \feh (SVM) \\
    18	 & $\sigma$(\feh) (SVM) \\
    19	 & $\teff$ (\ilium) \\
    20	 & $\sigma(\teff)$ (\ilium) \\
    21	 & $\A0$ (\ilium) \\
    22	 & $\sigma(\A0)$ (\ilium) \\
    23	 & $\logg$ (\ilium) \\
    24	 & $\sigma(\logg)$ (\ilium) \\
    25	 & \feh (\ilium) \\
    26	 & $\sigma$(\feh) (\ilium) \\
    27	 & $\teff$ (Aeneas p-model) \\
    28	 & $\sigma(\teff)$ (Aeneas p-model) \\
    29	 & $\A0$ (Aeneas p-model) \\
    30	 & $\sigma(\A0)$ (Aeneas p-model) \\
    31	 & $\logg$ (Aeneas p-model) \\
    32	 & $\sigma(\logg)$ (Aeneas p-model) \\
    33	 & \feh (Aeneas p-model) \\
    34	 & $\sigma$(\feh) (Aeneas p-model) \\
    35	 & $\teff$ (Aeneas pq-model) \\
    36	 & $\sigma(\teff)$ (Aeneas pq-model) \\
    37	 & $\A0$ (Aeneas pq-model) \\
    38	 & $\sigma(\A0)$ (Aeneas pq-model) \\
    39	 & $\logg$ (Aeneas pq-model) \\
    40	 & $\sigma(\logg)$ (Aeneas pq-model) \\
    41	 & \feh (Aeneas pq-model) \\
    42	 & $\sigma$(\feh) (Aeneas pq-model) \\
\hline
\end{tabular}
\end{center}
\end{table}

Tables~\ref{tab:datag15}, \ref{tab:datag19}, and \ref{tab:datagmixed} (available online) give the full results from the four methods on the three test data sets at G=15, G=19, and G=6--20 (mixed magnitudes), respectively. The tables have identical formats, with the columns defined in Table~\ref{tab:datacols}. 
SVM and \ilium\ return single values of the APs. Aeneas provides a full posterior PDF over the four APs: The AP estimate provided here is the mean of that PDF.
The precisions, $\sigma()$, are defined as follows. For both \ilium\ and Aeneas the precisions are 
estimates of the actual uncertainty in the AP estimate.
The \ilium\ precision is defined by equation 11 in CBJ10, which
shows how the covariance matrix of the APs relates to the covariance matrix of the BP/RP spectrum. The precisions are the square roots of the diagonal of this covariance matrix.
The Aeneas precision is half the interval which encloses 68\% of the posterior probability (equal to 1$\sigma$ if the distributions are Gaussian).
SVMs do not provide uncertainty estimates, and the ``precisions'' listed here are just averages of the residuals over small ranges of the AP space. They are, if you will, statistical precisions, as they do not take into account the specific input data. They should be used with caution.
Note that for \ilium\ the precisions appear to be overestimated, and for Aeneas they appear to be underestimated (thanks to Rene Andrae for identifying this). This may be a consequence of various approximations adopted in the algorithms, and is being investigated.

\label{lastpage}

\end{document}